\documentclass[10pt,journal,compsoc]{IEEEtran}

\ifCLASSOPTIONcompsoc
  \usepackage[nocompress]{cite}
\else
  \usepackage{cite}
\fi

\ifCLASSINFOpdf
 
\else
  
\fi

\usepackage{ragged2e}
\usepackage{amsmath}
\usepackage{amssymb}
\usepackage{algorithm}
\usepackage{algpseudocode}
\usepackage{multirow}
\usepackage{array}
\usepackage{graphicx}
\usepackage[table]{xcolor}
\usepackage{colortbl}
\usepackage{url}

\begin{document}
\title{Towards Robust and Generalizable Continuous Space-Time Video Super-Resolution with Events}

\author{Shuoyan~Wei,
        Feng~Li,
        Shengeng Tang,
        Runmin~Cong,~\IEEEmembership{Senior Member,~IEEE,}
        Yao~Zhao,~\IEEEmembership{Fellow,~IEEE,}
        Meng~Wang,~\IEEEmembership{Fellow,~IEEE,}
        and~Huihui~Bai,~\IEEEmembership{Senior Member,~IEEE}
\IEEEcompsocitemizethanks{\IEEEcompsocthanksitem S. Wei (shuoyan.wei@bjtu.edu.cn), Y. Zhao, and H. Bai are with the Institute of Information Science, Beijing Jiaotong University, Beijing 100044, China, and Beijing Key Laboratory of Advanced Information Science and Network Technology, Beijing 100044, China.
\IEEEcompsocthanksitem F. Li, S. Tang, and M. Wang are with Hefei University of Technology, Hefei 230009, China.
\IEEEcompsocthanksitem R. Cong is with Shandong University, Jinan 250100, China.
\IEEEcompsocthanksitem Corresponding author: Feng Li (fengli@hfut.edu.cn).}}

\markboth{Preprint, under review}%
{Wei \MakeLowercase{\textit{et al.}}: EvEnhancerPlus: Efficient and Generalizable Continuous Space-Time Video Super-Resolution with Events}

\IEEEtitleabstractindextext{%
\begin{abstract}
\justifying
Continuous space-time video super-resolution (C-STVSR) has garnered increasing interest for its capability to reconstruct high-resolution and high-frame-rate videos at arbitrary spatial and temporal scales. However, prevailing methods often generalize poorly, producing unsatisfactory results when applied to out-of-distribution (OOD) scales. To overcome this limitation, we present \textbf{EvEnhancer}, a novel approach that marries the unique properties of high temporal resolution and high dynamic range encapsulated in event streams to achieve robust and generalizable C-STVSR. Our approach incorporates event-adapted synthesis that capitalizes on the spatiotemporal correlations between frames and events to capture long-term motion trajectories, enabling adaptive interpolation and fusion across space and time. This is then coupled with a local implicit video transformer that integrates local implicit video neural function with cross-scale spatiotemporal attention to learn continuous video representations and generate plausible videos at arbitrary resolutions and frame rates. We further develop \textbf{EvEnhancerPlus}, which builds a controllable switching mechanism that dynamically determines the reconstruction difficulty for each spatiotemporal pixel based on local event statistics. This allows the model to adaptively route reconstruction along the most suitable pathways at a fine-grained pixel level, substantially reducing computational overhead while maintaining excellent performance. Furthermore, we devise a cross-derivative training strategy that stabilizes the convergence of such a multi-pathway framework through staged cross-optimization. Extensive experiments demonstrate that our method achieves state-of-the-art performance on both synthetic and real-world datasets, while maintaining superior generalizability at OOD scales. The code is available at \url{https://github.com/W-Shuoyan/EvEnhancerPlus}.

\end{abstract}

}

\maketitle

\IEEEdisplaynontitleabstractindextext

\IEEEpeerreviewmaketitle

\section{Introduction}
\label{sec:introduction}

\IEEEPARstart{V}{ideo} super-resolution (VSR) endeavors to recover high-resolution (HR) videos from their low-resolution (LR) counterparts, which is a fundamental task in low-level computer vision. However, in practice, video resources are mostly archived with both reduced spatial resolution and constrained frame rates (\textit{i.e.}, temporal resolution). Consequently, simultaneous restoration of spatial and temporal quality, \emph{i.e.}, generating HR and high-frame-rate (HFR) videos, poses significant challenges and remains an essential requirement for a wide range of applications~\cite{zhang2019two, wang2020dual,wang2023compression, yoo2023video,xu2024ibvc,li2024enhanced} and user experience~\cite{hou2022perceptual,rahimi2023spatio,wu2024perception}. 

A naive approach is to cascade separate VSR~\cite{wang2019edvr, li2019fast, haris2019recurrent, li2020learning, chan2021basicvsr, chan2022basicvsr++} and video frame interpolation (VFI)~\cite{niklaus2017video1, reda2019unsupervised, xu2019quadratic, lee2020adacof, chi2020all} or reverse workflow, which overlooks the correlations between the two sub-tasks, leading to suboptimal exploitation of spatiotemporal information in videos. Some methods have integrated VSR and VFI into a unified framework, \emph{i.e.}, space-time video super-resolution (STVSR)~\cite{haris2020space, xiang2020zooming, kim2020fisr, hu2022spatial, wang2022bi, wang2023stdan, hu2023cycmunet+}, which super-resolves the videos with intermediate frame interpolation to increase the spatial and temporal resolution in a single joint stage. Nevertheless, these methods are typically restricted to fixed discrete space and time magnifications, lacking support for arbitrary controls. To overcome this limitation, some researchers study continuous VFI~\cite{jiang2018super, bao2019depth, huang2022real, zhang2023extracting} or continuous VSR~\cite{lu2023learning, li2024savsr, huang2024arbitrary, shang2025arbitrary} methods, where the former is capable of variable frame interpolation while the latter enables arbitrary spatial upsampling. They operate in isolation and address only one aspect, either temporal or spatial continuity, failing to provide a holistic solution for continuous reconstruction across both dimensions simultaneously.

\begin{figure*}[t]
\centering
\includegraphics[width=\textwidth]{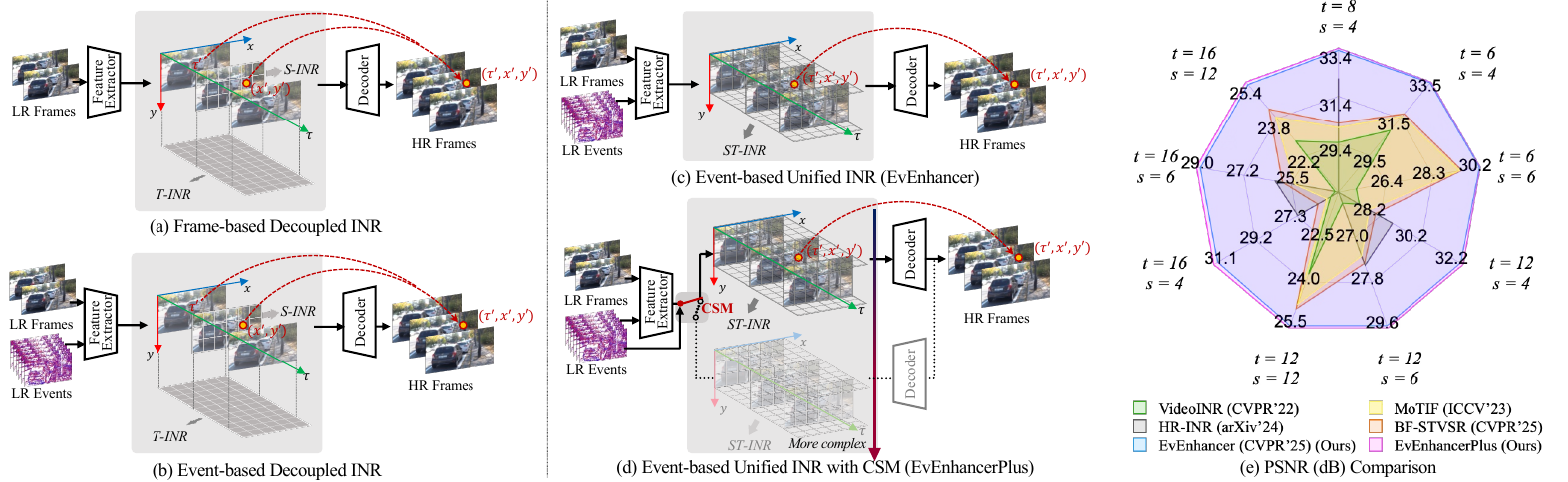}
\caption{Comparisons of different C-STVSR methods, including VideoINR~\cite{chen2022videoinr}, MoTIF~\cite{chen2023motif}, BF-STVSR~\cite{kim2025bf}, HR-INR~\cite{lu2024hr}, our \textbf{EvEnhancer} and \textbf{EvEnhancerPlus}. (a)-(d) illustrate the different methodology to implement INR, highlighted with a gray background. (e) illustrates the PSNR (dB) comparison for different spatial upsampling scale $s$ and temporal upsampling scale $t$ on GoPro~\cite{nah2017deep} among these methods.}
\label{fig:prf}
\end{figure*}
In recent years, continuous STVSR (C-STVSR)~\cite{chen2022videoinr, chen2023motif, lu2024hr, kim2025bf} has emerged as a promising approach, which learns video implicit neural representation (INR) to decode 3D spatiotemporal coordinates into RGB values, allowing for the generation of HR and HFR videos arbitrarily. However, as shown in Fig.~\ref{fig:prf}(a), current attempts~\cite{chen2022videoinr, chen2023motif, kim2025bf} mainly decouple the spatiotemporal INR (ST-INR) of video into separate spatial (S-INR) and temporal (T-INR), predicting continuous spatial and temporal feature domains independently. This design fails to fully capitalize on intrinsic spatiotemporal dependencies in the continuous domain, thus hindering reconstruction performance, particularly at scales outside the training distribution. In addition, 
these methods approximate intermediate motion fields between consecutive frames captured from conventional frame-based cameras, which proves insufficient for capturing accurate inter-frame temporal features under large or non-linear motions.

More recently, motivated by the exceptionally high temporal resolution and high dynamic range characteristics of event cameras, many works~\cite{yu2021training,han2021evintsr,jing2021turning,lu2023learning,lin2020learning,tulyakov2021time,he2022timereplayer,liu2025timetracker} introduce event data into video restoration tasks to fulfill inter-frame motion fields. Although promising improvements have been achieved, the potential of events in C-STVSR is largely unexplored. HR-INR~\cite{lu2024hr} leverages event temporal pyramid representation for dynamic motion, yet it remains constrained by the conventional paradigm of separate INR learning (see Fig.~\ref{fig:prf}(b)).

In light of the above discussion, we propose an innovative event-based approach \textbf{EvEnhancer}, which adeptly integrates the spatiotemporal synergies between video frames and event streams to learn continuous video representations for effective and generalized C-STVSR. Our approach first devises an event-adapted synthesis module (EASM) that executes event-modulated alignment and bidirectional recurrent compensation to explore the event information for long-term holistic motion trajectory modeling. This module incorporates motion cues from input frames with event-based modulation in both forward and backward temporal directions, enabling robust acquisition of latent inter-frame features. We then recurrently propagate the event stream across time and fuse it with aligned frame features in both directions to maximize the gathering of temporal information. In contrast to prior decoupling strategies~\cite{chen2022videoinr, chen2023motif, lu2024hr, kim2025bf} for video INR, we propose a local implicit video transformer (LIVT) that integrates local implicit video neural function with spatiotemporal attention. LIVT calculates cross-scale 3D attention based on the queried space-time coordinates in the continuous domain and the features derived from EASM within a local sampling grid. In this way, as shown in Fig.~\ref{fig:prf}(c), we can efficiently obtain an accurate unified video INR that directly maps any queried continuous spatiotemporal coordinate to its corresponding RGB value, consequently super-resolving videos in arbitrary frames and resolutions. 

Moreover, in C-STVSR, an intuitive observation is that the reconstruction difficulty of a pixel is strongly correlated with its spatiotemporal variation. Pixels in regions that exhibit significant motion or fine details often demand more sophisticated processing, while those in static or homogeneous areas can be reconstructed efficiently with simple operations. However, existing methods~\cite{chen2022videoinr,chen2023motif,kim2025bf,lu2024hr} process all pixels uniformly through identical spatial/temporal INRs for upsampling, which causes unnecessary computational overhead. Notably, event streams record the logarithmic intensity changes of each pixel, which provides a straightforward means of assessing such local reconstruction difficulty. To overcome this drawback, we formalize the theoretical relationship between event streams and latent frame changes, further presenting \textbf{EvEnhancerPlus} (see Fig.~\ref{fig:prf}(d)). It builds a parameter-free controllable switching mechanism (CSM) that dynamically determines the reconstruction difficulty for each spatiotemporal pixel based on local event statistics, enabling fine-grained adaptive routing of pixels to appropriately complex reconstruction pathways. This not only improves computational efficiency but also allows users to control the trade-offs between performance and efficiency during inference. To effectively integrate multiple pathways within a unified model, a cross-derivative training strategy is developed. We first train derivative networks for individual pathways independently and then integrate them into a single framework, which is finally fine-tuned using cross-initialization with weights inherited from pre-trained derivative networks. This ensures stable convergence and effective cooperation between pathways in \textbf{EvEnhancerPlus}. Experiments demonstrate that both \textbf{EvEnhancer} and \textbf{EvEnhancerPlus} achieve superior effectiveness in arbitrary spatial and temporal scales against existing state-of-the-art (SOTA) methods within training distribution (\textit{In-Dist.}). We also validate their preferable
generalizability at out-of-distribution (\textit{OOD}) scales (\textit{e.g.} Fig.~\ref{fig:prf}(e)). Furthermore, by leveraging CSM, \textbf{EvEnhancerPlus} reduces computational cost to about 85\% of that in \textbf{EvEnhancer} while maintaining excellent performance. Main contributions are as follows:
\begin{itemize}
    \item  We propose \textbf{EvEnhancer} that subtly marries the unique advantages of event streams with video frames for C-STVSR. Experiments validate its superior performance over recent SOTA methods.
    
    \item We propose ESAM that enables long-term holistic motion trajectory modeling to acquire informative spatiotemporal features.
    \item We propose LIVT that integrates local implicit video neural function with cross-scale spatiotemporal attention to learn an accurate unified video INR.
    \item We further improve \textbf{EvEnhancer} with a parameter-free CSM based on pixel-level reconstruction difficulty, \emph{i.e.}, \textbf{EvEnhancerPlus}, optimized by a cross-derivative training strategy, which achieves an optimal trade-off between efficiency and effectiveness. 
\end{itemize}

This paper is an extension of the conference version~\cite{wei2025evenhancer} published in CVPR 2025, referred to as \textbf{EvEnhancer}. Compared with the conference version, we significantly extend \textbf{EvEnhancer} to \textbf{EvEnhancerPlus} in the following aspects: 1) We formalize the theoretical relationship between event streams and latent frame changes. 2) We design the CSM, which is non-parametric and dynamically routes each pixel to computationally appropriate reconstruction pathways based on local event statistics, optimizing both efficiency and effectiveness. 3) We introduce a cross-derivative training strategy that can effectively stabilize the convergence of \textbf{EvEnhancerPlus}. 

\section{Related Work}
\label{sec:related}

\subsection{Video Super-Resolution (VSR)}
Conventional VSR aims to reconstruct HR videos from LR counterparts, which heavily relies on motion estimation and compensation between frames. Earlier methods address this problem by learning frame-level optical flows~\cite{li2020learning,wang2019learning}. Later methods apply deformable convolution \cite{wang2019edvr,tian2020tdan,chan2021basicvsr} or 3D convolution~\cite{huang2017video,isobe2020video} to conduct implicit feature-level motion compensation, exhibiting improved effectiveness and efficiency. Some methods are inspired by attention mechanisms~\cite{wang2018non,isobe2020revisiting}, incorporating spatial-temporal attention~\cite{yi2019progressive,li2020mucan,yu2022memory,zhou2022revisiting}, video transformers~\cite{liu2022learning,liang2022recurrent,qiu2023learning,zhou2024video}, or state space models~\cite{tran2025vsrm} to capture long-range spatiotemporal dependencies. Event cameras are famous as bio-inspired asynchronous sensors with high temporal resolution, high dynamic range, and low latency, which also have advanced VSR~\cite{han2021evintsr,jing2021turning,lu2023learning,xiao2024asymmetric,xiao2024event,kai2025event}. For example, E-VSR~\cite{jing2021turning} introduces the pioneering event-based VSR method that utilizes event-based asynchronous interpolation to synthesize asynchronous neighboring frames. EvTexture~\cite{kai2024evtexture} regards the voxelized events as high-frequency signals to improve the textures in RGB frames, achieving more effective VSR. Xiao~\emph{et al.}~\cite{xiao2025event} integrate event information in Mamba-based models to capture fine motion details. EGVSR~\cite{lu2023learning} unleashes the potential of events in arbitrary VSR, which learns spatial INR from both modalities but ignores the temporal flexibility. 

\subsection{Video Frame Interpolation (VFI)}
VFI involves interpolating latent frames between consecutive video frames to enhance the temporal resolution. Flow-based methods~\cite{xu2019quadratic, park2021asymmetric, kalluri2023flavr} leverage optical flow networks~\cite{dosovitskiy2015flownet,ilg2017flownet,sun2018pwc} to estimate the bidirectional flows between frames. Kernel-based methods \cite{niklaus2017video1, niklaus2017video2, cheng2020video} utilize local convolution to map inter-frame motion, which synthesizes target pixels by convolving over input frames. In addition, some recent works focus on splatting-based methods~\cite{niklaus2020softmax, niklaus2023splatting,hu2023video} or transformer-based methods~\cite{lu2022video, shi2022video, zhang2023extracting, liu2024sparse}, showing significant advances. As for event-based VFI, TimeLens \cite{tulyakov2021time} and TimeLens++ \cite{tulyakov2022time} innovatively integrate deformation and event-based synthesis processes. TimeLens-XL~\cite{ma2024timelens} utilizes the decomposition of long-term complex motion into multi-step small motions to improve efficiency. CBMNet \cite{kim2023event} utilizes cross-modality information to estimate inter-frame motion fields directly. REFID \cite{sun2023event}, as a flexible method, exhibits robustness even with blurred reference frames. Additionally, several generative methods~\cite{chen2025repurposing, zhang2025egvd} exploit the powerful priors of diffusion models to enhance the generalization in real-world data. However, similar to conventional VSR, these methods primarily focus on temporal upsampling independently, where few studies address a unified approach for both.

\subsection{Space-Time Video Super-Resolution (STVSR)}
STVSR can be seen as a combination of VSR and VFI, which reconstructs HR videos spatially and temporally. Most existing methods, including traditional~\cite{shechtman2005space,shahar2011space} and deep learning methods~\cite{xiang2020zooming, geng2022rstt, huang2024scale}, are prone to achieving STVSR with fixed spatial and temporal scales, denoted by F-STVSR. To mitigate the temporal scale constraints, TMNet \cite{xu2021temporal} proposes controlled frame interpolation with temporal modulation. SAFA~\cite{huang2024scale} introduces a scale-adaptive feature aggregation network that iteratively estimates motion with trainable block-wise scale selection. EvSTVSR~\cite{yan2025evstvsr} combines frame-based optical flow prediction and event-based dense temporal information, coupled with implicit sampling to solve the STVSR task. Indeed, while these methods introduce temporal flexibility to some extent, they remain constrained to fixed or discrete spatial upsampling factors.

\subsection{Implicit Neural Representation (INR) in STVSR}
As a method of continuous representation of signals, INR is a common means of continuous super-resolution, including image \cite{chen2021learning, yang2021implicit, lee2022local, chen2023cascaded, he2024latent} and video \cite{chen2022videoinr, lu2023learning, chen2023motif, lu2024hr, yan2025evstvsr} tasks. LIIF~\cite{chen2021learning} represents the first effort that learns the local implicit image function, enabling continuous extrapolation for arbitrary-scale super-resolution. Inspired by LIIF \cite{chen2021learning}, VideoINR \cite{chen2022videoinr} pioneers C-STVSR by learning separate spatial and temporal INRs. MoTIF \cite{chen2023motif} adds forward spatiotemporal INR to enhance motion field estimation. BF-STVSR~\cite{kim2025bf} applies a B-spline mapper to tackle temporal interpolation and a Fourier mapper to capture spatial frequencies sequentially. HR-INR~\cite{lu2024hr} adheres to the same decoupling paradigm as VideoINR~\cite{chen2022videoinr}, which combines holistic event-frame feature extraction with INR decoding to realize C-STVSR, but requires multiple input frames to improve details, inadvertently introducing error accumulation and computational redundancy. Unlike existing methods, we break the decoupling paradigm. We propose the EASM 
that performs event-modulated alignment and bidirectional recurrent compensation to model long-term motion trajectories. Upon this, our LIVT leverages high temporal resolution grids derived from event data to learn a unified spatiotemporal INR, achieving better generalization with reduced model complexity.

\begin{figure*}[t]
\centering
\includegraphics[width=\textwidth]{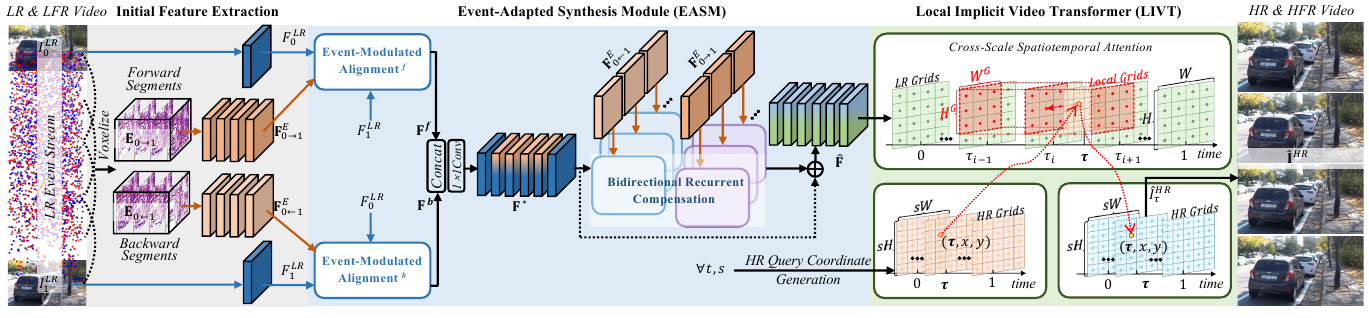}
\caption{
The overall backbone of our \textbf{EvEnhancer} and \textbf{EvEnhancerPlus} models consists of an event-adapted synthesis module (EASM), and a local implicit video transformer (LIVT).
}
\label{fig:network}
\end{figure*}

\section{Proposed Method}
Given LR, low-frame-rate (LFR) videos and the corresponding event streams as input, we present an event-based C-STVSR approach named \textbf{EvEnhancer} to reconstruct HR and HFR videos on arbitrary spatial and temporal scales. Before elaborating on the components of our methodology, we first provide the details for the event representation in this work.

\subsection{Event Representation}
Event cameras asynchronously capture the logarithmic intensity change $\Delta L$ of each pixel at coordinate $(x,y)$ within the intensity image $I$. When $\Delta L_{\tau - \Delta \tau \rightarrow \tau}(x, y)=\log I(\tau, x, y) - \log I(\tau - \Delta \tau, x, y)$ in the duration $[\tau - \Delta \tau, \tau]$ surpasses the contrast threshold $c$, an event $r = (\tau, x, y, p)$ is triggered at timestamp $\tau$, where $p = \pm 1$ is the polarity. This process can be described as
\begin{equation}
\label{eq:1}
    p = 
    \begin{cases}
    +1, & \Delta L(x, y) > +c, \\
    -1, & \Delta L(x, y) < -c.
    \end{cases}
\end{equation}
For any timestamps $\tau_s, \tau_e$, the logarithmic intensity change $\Delta L_{\tau_s \rightarrow \tau_e}$ between frames $I_{\tau_s}$ and $I_{\tau_e}$ is derived as
\begin{equation}
\label{eq:2}
    \Delta L_{\tau_s \rightarrow \tau_e}(x,y) = c \int_{\tau_s}^{\tau_e} p(\tau,x,y)\, d\tau.
\end{equation}

Compared to conventional RGB frames, a raw event stream constituted by large amounts of sparse points is capable of continuously capturing intensity changes with high temporal resolution. However, processing such an unstructured sparse format in deep networks is non-trivial. Following existing event-based methods~\cite{sun2023event}, we represent the event stream $\mathcal{E}_{\tau_s \rightarrow \tau_e}$ within the duration $[\tau_s, \tau_e]$ into a spatiotemporal voxel grid $V_{\tau_s \rightarrow \tau_e}\in\mathbb{R}^{H\times W\times (M+1)}$ linearly, where $H\times W$ denotes the spatial size, aligned with the RGB frame. $(M+1)$ is the number of time bins. The $m$-th time bin of the voxel grid acquires the distribution from each event $r$ as
\begin{equation}
\label{eq:3}
    V_m=\sum_{r}p_r\max(0, 1-|(m-1)-\frac{\tau_{r}- \tau_s}{\tau_e - \tau_s}M)|),
\end{equation}
where $m\in\{1, 2, ..., (M+1)\}$. $p_r$ and $\tau_r$ are the polarity and timestamp of the event $r$, respectively.
To mitigate the impact of hot pixels~\cite{zhu2019unsupervised}, we normalize the voxel grid $V$ as
\begin{equation}
\label{eq:3-1}
    \overline{V}_m=\frac{\min(V_m, \eta)}{\eta},
\end{equation}
where $\eta$ is the $98$-th percentile value in the non-zero values of $V$. To utilize continuous temporal information~\cite{sun2023event}, we combine two consecutive time bins into one segment, resulting in $M$ segments $[E_1, E_2,..., E_m,..., E_{M-1}, E_M]$. Each segment $E_{m}$ records the intensity information within a time window nearest to $\tau_m$ (see Fig.~\ref{fig:network}). 

\subsection{Overview}
Let $\mathbf{I}^{\textit{LR}}=\{I_0^{\textit{LR}}, I_1^{\textit{LR}}\}\in\mathbb{R}^{H\times W\times 3}$ denote two LR RGB frames and $\mathbf{E}_{0 \rightarrow 1}$ denote the event segments from their corresponding event stream $\mathcal{E}_{0 \rightarrow 1}$. We propose \textbf{EvEnhancer} that super-resolves $\mathbf{I}^{\textit{LR}}$ to $(t+1)$ HR video frames $\hat{\mathbf{I}}^{\textit{HR}}=\{\hat{I}_{\tau}^{\textit{HR}}\}\in\mathbb{R}^{sH\times sW\times 3}$ with any spatial scale $s\geq1$ and temporal scale $t\geq1$. The overall network is illustrated in Fig.~\ref{fig:network}, which consists of an event-adapted synthesis module (EASM) and a local implicit video transformer (LIVT). 

Generally, we first extract initial frame features $F_0^{\textit{LR}}$ and $F_1^{\textit{LR}}$. Then, we use the same structure to extract the initial features of event segments $\mathbf{F}_{0 \rightarrow 1}^{E}=\{F^{E,m}_{0\rightarrow 1}\}^M_{m=1}$. Considering that the latent RGB inter-frame features can be captured in forward and backward directions, we also reverse the time and polarity of $\mathcal{E}_{0 \rightarrow 1}$ to derive backward event features $\mathbf{F}^E_{0 \leftarrow 1}$. Then, EASM models long-term holistic motion trajectories guided by events to generate a spatiotemporal feature sequence $\hat{\mathbf{F}}$ with high temporal resolution. Finally, LIVT integrates cross-scale spatiotemporal attention with local implicit function to learn continuous video INR that decodes arbitrary space-time coordinates into RGB values, recovering HR video frames $\hat{\mathbf{I}}^{\textit{HR}}$ with HFR.

\subsection{Event-Adapted Synthesis Module}
\begin{figure*}[t]
\centering
\includegraphics[width=\textwidth]{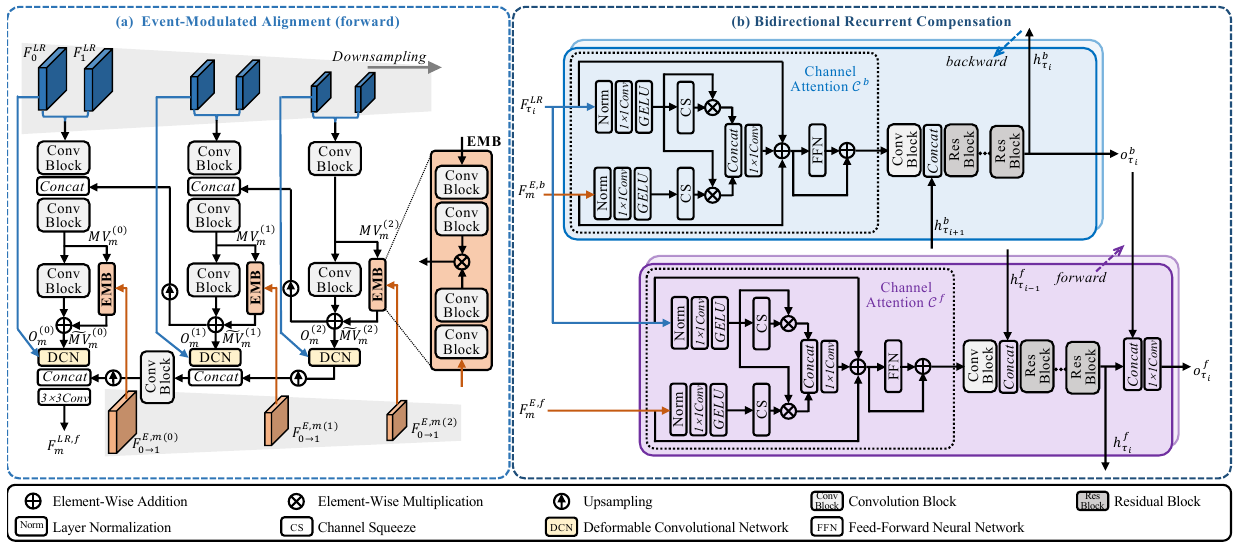}
\caption{
The detail architecture of the event-adapted synthesis module (EASM), which contains two steps: (a) event-modulated alignment, and (b) bidirectional recurrent compensation. ``EMB'': event modulation block.
}
\label{fig:EASM}
\end{figure*}

As depicted in Fig.~\ref{fig:EASM}, EASM contains two steps: 1) Event-modulated alignment (EMA) incorporates motion cues between input frames with event modulation to acquire latent inter-frame features. Based on the event features and reverse ones, we deploy such alignment in parallel to model the motions forward (``$f$'') and backward (``$b$''). 2) Bidirectional recurrent compensation (BRC) propagates the event stream across time and fuses it with acquired features in both directions to maximize the gathering of temporal information.

\subsubsection{Event-Modulated Alignment}
Given input frame features $F^{\textit{LR}}_0$ and $F^{\textit{LR}}_1$, our objective is to infer their intermediate frame features according to $M$ event segment features. Motivated by the proven efficacy of the PCD module in video restoration~\cite{wang2019edvr, xiang2020zooming}, we construct a customized pyramid integrated with event modulation to learn motion offsets and feature alignment progressively. For clarity, we describe the forward alignment process as an illustrative example.

Specifically, for the forward alignment of $m$-th segment, as shown in Fig.~\ref{fig:EASM}(a), we implement a 3-level pyramid by downsampling the frame and forward event features ($F^{\textit{LR}}_0$, $F^{\textit{LR}}_1$, $\mathbf{F}_{0\rightarrow 1}^{E}$) to 3 different spatial scales \(\times 1\), \(\times \frac{1}{2}\), \(\times \frac{1}{4}\) via stride convolutions. At each level $l$, we first calculate an initial motion vector by a convolutional block on concatenated frame features, which is then fused with the learned offset $O_{m}^{(l+1)}$ from the $(l+1)$-th level to obtain a motion vector ${MV}_{m}^{(l)}$ nearest to the $m$-th segment. Then, we design an event modulation block (EMB) $\mathcal{M}^{(l)}(\cdot)$ to generate modulated $\widetilde{MV}_{m}^{(l)}$ controlled by the event feature $F^{E,m}_{0\rightarrow1}$. Compared to previous PCD-based methods~\cite{wang2019edvr,xu2021temporal,xiang2020zooming}, our event-based approach can furnish more deterministic motion cues to facilitate frame synthesis, formulated as
\begin{equation}
\label{eq:4}
    \widetilde{MV}_{m}^{(l)} = \mathcal{M}^{(l)}({MV}_{m}^{(l)}, F^{E, m(l)}_{0\rightarrow1}),
\end{equation}
where $F^{E, m(l)}_{0\rightarrow1}$ denotes the downsampled event feature at level $l$. Then, we can learn the offset ${O}_{m}^{(l)}$ from ${MV}_{m}^{(l)}$ and $\widetilde{MV}_{m}^{(l)}$. Following~\cite{wang2019edvr}, we use the deformable convolutional network (DCN)~\cite{zhu2019deformable} for feature alignment and interpolation. The deformable alignments at different levels are cascaded and aggregated coarse-to-fine for accuracy improvement via the pyramid workflow. 

Since there are $M$ event features $\{F^{E,m}_{0\rightarrow1}\}^M_{m=1}$, through the forward modulated alignment, we can produce $M$ intermediate frame features $\mathbf{F}^f=\{F^{LR,f}_{m}\}^M_{m=1}$. Similarly, for the backward modulated alignment, we can obtain $\mathbf{F}^b=\{F^{LR,b}_{m}\}^M_{m=1}$. These two are integrated by channel concatenation with a $1 \times 1$ convolution, forming the interpolated features $\mathbf{F}^{*}=\{F^{\textit{LR}}_{\tau_i}\}$ with the input frame features $F^{\textit{LR}}_0$, $F^{\textit{LR}}_1$, where the timestamp $\tau_i \in \{0, \frac{1}{M+1},....,\frac{m}{M+1},..., \frac{M-1}{M+1},\frac{M}{M+1},1\}$ is aligned with the input frames and $M$ event segments.

\subsubsection{Bidirectional Recurrent Compensation}
Given that the information encapsulated within each event segment is confined to a narrow time window, after the EMA, we propose bidirectional recurrent compensation (BRC) that iteratively fuses the event and frame features to model the global temporal information inherent in two modalities, as shown in Fig.~\ref{fig:EASM}(b). In each recurrent iteration, the compensation block receives backward and forward event segment features $F^{E,b}_{m}$, $F^{E,f}_{m}$ from the current timestamp $\tau_m$, the forward hidden state from $h^f_{\tau_{i-1}}$ the previous timestamp $\tau_{i-1}$, and the backward hidden state $h^b_{\tau_{i+1}}$ from the next timestamp $\tau_{i+1}$, and the aligned frame feature $F^{\textit{LR}}_{\tau_i}\in\mathbf{F^{*}}$. Hence, the recurrent compensation processes $\mathcal{R}^b(\cdot)$ and $\mathcal{R}^f(\cdot)$ at the timestamp $\tau_{i}$ are formulated as
\begin{equation}
\label{eq:5}
    \begin{aligned}
    o_{\tau_{i}}^{b}, h_{\tau_{i}}^{b} &= \mathcal{R}^b(\mathcal{C}^b(F^{\textit{LR}}_{\tau_{i}}, F^{E,b}_{m}), h_{\tau_{i+1}}^{b}), \\
    o_{\tau_{i}}^{f}, h_{\tau_{i}}^{f} &= \mathcal{R}^{f}(\mathcal{C}^f(F^{\textit{LR}}_{\tau_{i}}, F^{E,f}_{m}), h_{\tau_{i-1}}^{f}, o_{\tau_{i}}^{b}),\\
    \end{aligned}
\end{equation}
where $o_{\tau_{i}}^{b}$ and $h_{\tau_{i}}^{b}$ denote the output and hidden states respectively in the backward direction, $o_{\tau_{i}}^{f}$ and $h_{\tau_{i}}^{f}$ correspond to the forward. $\mathcal{C}^b$ and $\mathcal{C}^f$ calculate the channel attention on event and frame features during the forward and backward recurrent iterations, respectively. At sequence boundaries, if there is no preceding state or aligned event segments, the corresponding computation is skipped. By recurrently propagating the event information across time, we can comprehensively exploit the high temporal resolution properties of events to refine the frame features and maximize the gathering of temporal information. To retain valid information, a bidirectional feature fusion is conducted by the element-wise addition between the input feature fed into BRC and its output bidirectional features as
\begin{equation}
\label{eq:5-1}
    \hat{\mathbf{F}} = \mathbf{F^{*}} + \text{BRC}(\mathbf{F^{*}}).
\end{equation}

\subsection{Local Implicit Video Transformer}
After the EASM, we obtain the discrete latent feature sequence $\hat{\mathbf{F}}$. The other key in this method is to learn a continuous video INR that specifies the feature at the queried space-time coordinate $(\tau,x,y)$ is transferred to any target spatial scale $s$ and temporal scale $t$ (see the right portion of Fig.~\ref{fig:network}). Previous C-STVSR methods~\cite{chen2022videoinr, chen2023motif, lu2024hr, kim2025bf} learn spatial and temporal INR separately to realize video INR, which fails to fully capitalize on the spatiotemporal dependencies in the continuous domain. In this work, we propose LIVT, which calculates cross-scale 3D attention based on the queried space-time coordinates in the continuous domain over a local sampling grid to learn an accurate unified video INR directly.

\begin{figure}[t]
\centering
\includegraphics[width=\columnwidth]{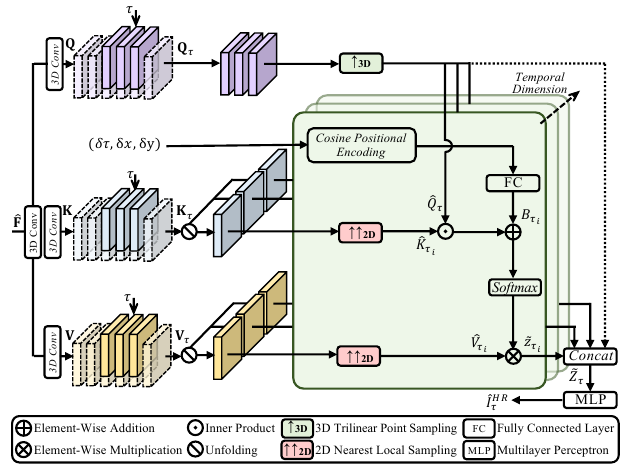}
\caption{
Structure of the local implicit video transformer (LIVT), which integrates 3D local spatiotemporal attention with implicit neural function to learn continuous video INR to reconstruct HR and HFR video frames.
}
\label{fig:LIVT}
\end{figure}

As shown in Fig.~\ref{fig:LIVT}, with the LR feature sequence $\hat{\mathbf{F}}$ from EASM, we encode them into three embedding spaces through 3D convolution: query $\mathbf{Q}$, key $\mathbf{K}$, and value $\mathbf{V}$. Notably, we do not calculate the spatiotemporal attention across the overall duration. Assuming the temporal length and spatial size of the local grid in LR scale are $T^{G}$ and $H^{G}\times W^{G}$, respectively. For the full duration $[0,1]$, to generate a frame at the target timestamp $\tau\in[0,1]$, we only need to select the $T^G$ corresponding features nearest to $\tau$ along the temporal dimension. 
We denote the set of candidate timestamps $\tau_i$ as $\mathbf{S}_{\tau} \subseteq [0, \frac{1}{M+1},....,\frac{M}{M+1},1]$, and the size of the $\lvert \mathbf{S}_{\tau} \rvert=T^G$, which is selected as follows: 
\begin{equation}
\label{eq:6}
     \mathbf{S}_{\tau} = \underset{\mathbf{S}_{\tau}}{\arg \min}\sum_{\tau_{i} \in \mathbf{S}_{\tau}} \lvert \tau_{i} - \tau \rvert.
\end{equation}
Based on Eq. (\ref{eq:6}), we conduct temporal selection on $\mathbf{Q}$, $\mathbf{K}$, and $\mathbf{V}$, and then reshape them to get $\mathbf{Q}_{\tau}$, $\mathbf{K}_{\tau}$, and $\mathbf{V}_{\tau}$, which allows the learning for 3D local attention in a sub-duration nearest to the target timestamp $\tau$. On top of this, we employ 3D trilinear sampling on $\mathbf{Q}_{\tau}$ to get the query $\widehat{Q}_{\tau} \in \mathbb{R}^{s^{2}HW \times 1 \times C}$ for a certain frame, where $C$ is the channel numbers. Besides, we employ nearest sampling on $\mathbf{K}_{\tau}$ and $\mathbf{V}_{\tau}$ to get $\widehat{K}_{\tau}$ and $\widehat{V}_{\tau}$, respectively. However, since both $\widehat{K}_{\tau}$ and $\widehat{V}_{\tau}$ have the size of $s^{2}HW \times T^{G}H^{G}W^{G} \times C$,  directly computing the attention would incur prohibitively high computational costs. To mitigate this, we propose to unfold the sequence along the temporal dimension and calculate cross-scale attention separately on $T^{G}$ embeddings. Thus, we sample from $\mathbf{K}_{\tau}$ and $\mathbf{V}_{\tau}$ using 2D nearest local sampling at each time slice $\tau_i$, getting $\widehat{K}_{\tau_{i}}$ and $\widehat{V}_{\tau_{i}}$, respectively. Moreover, we encode and reshape the spatiotemporal relative coordinates $\delta_\mathbf{C} = \{(\delta \tau, \delta x, \delta y)\}\in(-1,1)$ from each query point to all pixel points within its local grid, obtaining the spatiotemporal bias $B_{\tau_{i}}\in\mathbb{R}^{s^2HW\times H^G W^G\times3}$. The cosine positional encoding $\mathcal{G}(\cdot)$ used here is an extension of ~\cite{chen2023cascaded}
\begin{equation}
    \begin{aligned}
    \mathcal{G}(\delta_\mathbf{C})=[&\sin(2^0\delta_\mathbf{C}),\cos(2^0\delta_\mathbf{C}),...,\\
    &\sin(2^{L-1}\delta_\mathbf{C}),\cos(2^{L-1}\delta_\mathbf{C})],
    \end{aligned}
    \label{eq:7}
\end{equation}
where $L$ is a hyperparameter set to 10. 

We compute the inner product between $\widehat{Q}_{\tau}$ and $\widehat{K}_{\tau_{i}}$ and add it with $B_{\tau_{i}}$ to obtain the local attention map at the timestamp $\tau_{i}$ after a Softmax operation. After the matrix multiplication with $\widehat{V}_{\tau_{i}}$, we can derive the output embedding $\hat{z}_{\tau_{i}}$. This procedure is described as
\begin{equation}
\label{eq:8}
    \begin{aligned}
    \hat{z}_{\tau_{i}} = \text{Softmax}(\frac{\widehat{Q}_{\tau}\widehat{K}_{\tau_{i}}^{\top}}{\sqrt{C}}+B_{\tau_{i}}) \otimes \widehat{V}_{\tau_{i}}.
    \end{aligned}
\end{equation}

We concatenate the output embeddings $\hat{z}_{\tau_{i}}$ along the channel dimension to construct the complete spatiotemporal INR to yield the features $\hat{Z}_{\tau}$ corresponding to space-time coordinates. Following \cite{chen2021learning, chen2023cascaded, chen2022videoinr}, we decode the features into RGB values using a 5-layer MLP with the previous query $\widehat{Q}_{\tau}$ to reconstruct the target RGB frame $\hat{I}_\mathcal{\tau}^{\textit{HR}}$.

\begin{figure}[t]
\centering
\includegraphics[width=\columnwidth]{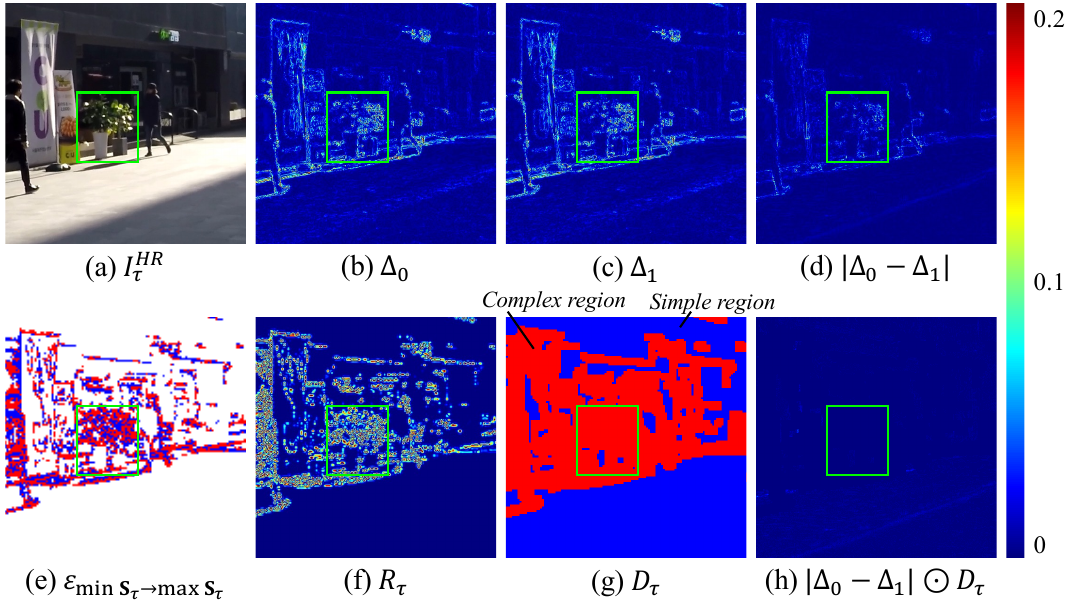}
\caption{The motivation of CSM. (a) is the ground-truth (GT) frame. (b) and (c) are the residual intensity maps that represent the difference between the GT frame and reconstructed frames by a simple upsampler $\mathcal{U}_0$ and a complex upsampler $\mathcal{U}_1$, respectively. (d) denotes the quality bias between (b) and (c). We use Eq. (\ref{eq:14}) to calculate the reconstruction difficulty of pixels (f) based on corresponding events (e). Based on Eq. (\ref{eq:11}), we can deploy the distributor $\mathcal{D}(\cdot)$ that derives the distribution map (g) from (f). The result (h) of the Hadamard product ``$\bigodot$'' between (d) and (g) reflects the reconstruction discrepancy between the remaining regions by \textbf{EvEnhancerPlus}.}
\label{fig:demonstration}
\end{figure}

\begin{figure*}[t]
\centering
\includegraphics[width=\textwidth]{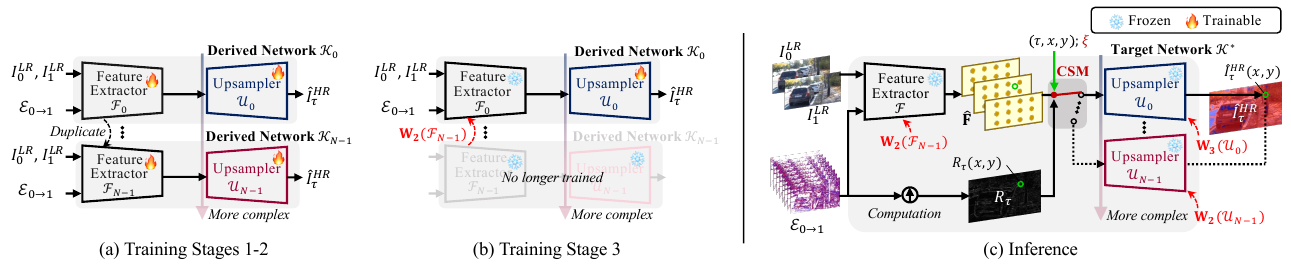}
\caption{The overall training and inference process of our \textbf{EvEnhancerPlus}. (a) and (b) show the training stages, which follow the cross-derivative training strategy. (c) shows the inference stage. The light gray background denotes each independent reconstruction network, and the red symbols indicate the weight used for a component during initialization.}
\label{fig:framework}
\end{figure*}

\subsection{EvEnhancerPlus}
As discussed, INR plays a significant role in C-STVSR to increase the spatial and temporal resolution of input videos. Here, we denote the module learning INR as an upsampler $\mathcal{U}(\cdot)$ that super-resolves the features $\hat{\mathbf{F}}$ to generate a pixel $\hat{I}_{\tau}(x,y)$ given the spatiotemporal coordinate $(\tau, x, y)$, can be formulated as 
\begin{equation}
\label{eq:9}
    \hat{I}_{\tau}^{\textit{HR}}(x,y)=\mathcal{U}(\hat{\mathbf{F}}\text{; }(\tau,x,y)).
\end{equation}

However, pixels in regions characterized by significant motion or intricate details necessitate sophisticated processing, whereas those in static or homogeneous regions can be efficiently reconstructed using simpler operations. Existing methods~\cite{chen2022videoinr, chen2023motif, lu2024hr, kim2025bf} employ a uniform pattern for all pixels, which can result in substantial computational redundancy, particularly for pixels with low reconstruction difficulty. As illustrated in Fig.~\ref{fig:demonstration}, we construct two upsampler examples, $\mathcal{U}_0$ and $\mathcal{U}_1$, by configuring LIVT with different channel numbers (16 and 64, respectively). It can be observed that for a given HR clean frame ${I}_{\tau}^{\textit{HR}}$, the simple upsampler $\mathcal{U}_0$ can achieve comparable reconstruction quality to the complex one $\mathcal{U}_1$ in certain regions (Fig.~\ref{fig:demonstration}(b) and \ref{fig:demonstration}(c)). The performance gap only arises in regions where $\mathcal{U}_0$ struggles, as evidenced by the difference $\vert \Delta_0$-$\Delta_1 \vert$ between the residual error maps $\Delta_0$, $\Delta_1$ (Fig.~\ref{fig:demonstration}(d)). This observation suggests that we can design a more efficient approach to selectively upsample pixels without compromising overall reconstruction performance.

In our event-based framework, as formulated in Eq. (\ref{eq:2}), there is a theoretical approximate relation between events and latent frame changes, \emph{i.e.}, the event streams inherently capture logarithmic variations per pixel. This property enables direct estimation of the spatiotemporal reconstruction difficulty for pixels using local event statistics. Therefore, we propose \textbf{EvEnhancerPlus}, which distributes different pixels to $N$ upsamplers $\{\mathcal{U}_{n}(\cdot)\}_{n=0}^{N-1}$ with different computations based on the event statistics, achieving fine-grained adaptive routing of pixels to appropriately complex reconstruction pathways. This can be expressed as    
\begin{equation}
\label{eq:10}
    \hat{I}_{\tau}^{\textit{HR}}(x,y)=\sum_{n=0}^{N-1}\mathcal{U}_{n}(\hat{\mathbf{F}}\text{; }(\tau,x,y)) \times \mathcal{D}_{n}(\tau,x,y),
\end{equation}
where $\mathcal{D}_{n}(\cdot)$ indicates the distributor formulated as
\begin{equation}
\label{eq:11}
    \mathcal{D}_{n}(\tau,x,y)=
    \begin{cases}
    1, & \text{if }R_\tau(x,y) \in (\xi_{n}, \xi_{n+1}], \\
    0, & \text{otherwise}.
    \end{cases}
\end{equation}
Here, $R_\tau(x,y)\in[0,1]$ is the normalized reconstruction difficulty based on event statistics. 
The boundary values of the reconstruction difficulty are defined as $\xi_0 = 0$ and $\xi_N = 1$, while the intermediate terms $\{\xi_n\}_{n=1}^{N-1} \in [0, 1]$ denote a set of adjustable trigger thresholds that are arranged in ascending order.

\subsubsection{Controllable Switch Mechanism}
Based on Eq. (\ref{eq:2}), for any duration $[\tau_s, \tau_e]$, the amount of logarithmic intensity change $|\Delta L_{\tau_s \rightarrow \tau_e}|$ can be calculated as
\begin{equation}
\label{eq:12}
    |\Delta L_{\tau_s \rightarrow \tau_e}| = c\ | \int_{\tau_s}^{\tau_e} p_r\, dr|.
\end{equation}
As shown in Fig.~\ref{fig:framework}(c), we propose the controllable switch mechanism (CSM) that serves as the distributor $\mathcal{D}_{j}(\cdot)$ to derive the reconstruction difficulty $R_\tau(x,y)$ from events. First, we employ the max-min normalization on $|\Delta L_{\tau_s \rightarrow \tau_e}|$ to eliminate the effect of the threshold $c$, formulated as
\begin{equation}
\label{eq:13}
    \Delta \overline{L}_{\tau_s \rightarrow \tau_e} = \frac{|\Delta L_{\tau_s \rightarrow \tau_e}|-\min{|\Delta L_{\tau_s \rightarrow \tau_e}|}}{\max{|\Delta L_{\tau_s \rightarrow \tau_e}|}-\min{|\Delta L_{\tau_s \rightarrow \tau_e}|}}.
\end{equation}

Considering the lower spatial resolution of event streams than the target RGB frame $\hat{I}_{\tau}^{\textit{HR}}$, we upsample $\Delta \overline{L}_{\tau_s \rightarrow \tau_e}$ to align with the spatial size of $\hat{I}_{\tau}^{\textit{HR}}$. Besides, in the upsampling process indicated in Eq. (\ref{eq:6}) and (\ref{eq:10}) in our LIVT, an output pixel located at the timestamp $\tau$ corresponds to the latent features at the timestamps $\tau_i \in \mathbf{S}_\tau$, where the intensity change is located at the duration $[\min{\mathbf{S}_\tau}, \max{\mathbf{S}_\tau}]$. Thus, we can obtain the reconstruction difficulty $R_\tau$ from $\Delta \overline{L}$:
\begin{equation}
\label{eq:14}
    R_\tau = \text{Up}(\Delta \overline{L}_{\min{\mathbf{S}_\tau} \rightarrow \max{\mathbf{S}_\tau}}),
\end{equation}
where $\text{Up}(\cdot)$ denotes the spatial upsampling, and $\Delta \overline{L}_{\min{\mathbf{S}_\tau} \rightarrow \max{\mathbf{S}_\tau}}$ is computed from Eq. (\ref{eq:13}), where the value of $|\Delta L_{\tau_s \rightarrow \tau_e}|$ becomes equivalent to $|\sum_{r}p_r|$ from events $\mathcal{E}_{\min{\mathbf{S}_\tau} \rightarrow \max{\mathbf{S}_\tau}}$ when $c$ is omitted.

As shown in Fig.~\ref{fig:demonstration}(e)-\ref{fig:demonstration}(g), assuming that there are two upsamplers ($N=2$), the distributor $\mathcal{D}(\cdot)$ using CSM can automatically split pixels into two types of difficulty based on $R_{\tau}$. In this way, our \textbf{EvEnhancerPlus} can achieve a highly competitive performance with the computationally expensive alternative of uniformly processing all pixels with the same complex upsampler (Fig.~\ref{fig:demonstration}(h) reveals the minimal reconstruction discrepancy under our method). Moreover, according to Eq. (\ref{eq:11}) and (\ref{eq:14}), CSM is parameter-free, which allows users to flexibly regulate the trade-off between performance and computation by controlling the trigger threshold $\xi_n$ during the inference process.

\subsubsection{Cross-Derivative Training Strategy}
As analyzed above, our \textbf{EvEnhancerPlus} involves the implementation of multiple upsamplers. If we train them simultaneously from scratch, the training process would be unstable. To alleviate such instability, we propose a cross-derivative training strategy that stabilizes convergence toward the target model weight $\mathbf{W}^*$. For clarity, we simplify the overall network $\mathcal{K}^*$ into a composite feature extractor $\mathcal{F}$ and $N$ upsamplers $\{\mathcal{U}_n\}_{n=0}^{N-1}$ ordered from the smallest to largest by their computational complexity.

To be specific, as shown in Fig.~\ref{fig:framework}(a) and \ref{fig:framework}(b), there are three stages for training \textbf{EvEnhancerPlus}. In Stage 1, both the temporal scale factor $t$ and the spatial scale factor $s$ are fixed. $\mathcal{F}$ is spliced with different $\mathcal{U}_n$ deriving $N$ independent networks $\{\mathcal{K}_n\}_{n=0}^{N-1}$. For each $\mathcal{K}_n$, the extractor weight $\mathbf{W}_1(\mathcal{F}_n)$ and upsampler weight $\mathbf{W}_1(\mathcal{U}_n)$ are randomly initialized and updated during training. In Stage 2, the temporal scale factor $t$ remains fixed while the spatial scale factor $s$ varies uniformly. The pre-trained weights $\mathbf{W}_1(\mathcal{F}_n)$ and $\mathbf{W}_1(\mathcal{U}_n)$ from Stage 1 are used to initialize corresponding networks, which are further fine-tuned under varying spatial scales to obtain $\mathbf{W}_2(\mathcal{F}_n)$ and $\mathbf{W}_2(\mathcal{U}_n)$. To ensure robust handling of pixels with high reconstruction difficulty, following the order, the weights derived from the most complex upsampler, $\mathbf{W}_2(\mathcal{F}_{N-1})$ and $\mathbf{W}_2(\mathcal{U}_{N-1})$, are determined as the target weights for the full network $\mathcal{K}^*$, denoted as $\mathbf{W}^*(\mathcal{F})$ and $\mathbf{W}^*(\mathcal{U}_{N-1})$. As for Stage 3, the scale factor settings are consistent with those in Stage 2, and the weight of the feature extractor is frozen to $\mathbf{W}^*(\mathcal{F})$. Under this configuration, only the remaining upsampler weights are updated based on $\mathbf{W}_2(\mathcal{U}_{n})$, producing the optimized weights $\{\mathbf{W}^*(\mathcal{U}_{n})\}_{n=0}^{N-2}$ for each upsampling pathway. 

In \textbf{EvEnhancerPlus}, the original LIVT from \textbf{EvEnhancer} is retained as the most complex upsampler, while the rest are designed to be more lightweight. Therefore, the full model can ensure optimal computation allocation by adaptively routing pixels to different upsamplers. The details of our cross-derivative training strategy are in Algorithm~\ref{alg:1}.

\begin{algorithm}[t]
\scriptsize
\caption{Cross-Derivative Training Strategy}
\label{alg:1}
\begin{algorithmic}[1]
\Statex \hspace{-2em} \textbf{Input:} \textbf{EvEnhancerPlus} network $\mathcal{K}^* \gets [\mathcal{F};\{\mathcal{U}_n\}_{n=0}^{N-1}]$; number of iterations $d_1,d_2,d_3$; and temporal scale factor $t$ and spatial scale factor $s$
\State \textbf{\textit{Stage 1:}} Fix $t$ and $s$
\For{$n = 0$ to $N-1$}
    \State Duplicate $\mathcal{F}_n \gets \mathcal{F}$
    \State Splice $\mathcal{F}_n$ and $\mathcal{U}_n$ into a derived network $\mathcal{K}_n\gets [\mathcal{F}_n;\mathcal{U}_n]$
    \State Initialize weights $\mathbf{W}_1(\mathcal{F}_n)$, $\mathbf{W}_1(\mathcal{U}_n)$
    \For{$j = 0$ to $d_1-1$}
        \State Update all weights $\mathbf{W}_1(\mathcal{F}_n) \gets \mathbf{W}_1(\mathcal{F}_{n})$, $\mathbf{W}_1(\mathcal{U}_n) \gets \mathbf{W}_1(\mathcal{U}_{n})$
    \EndFor
\EndFor
\State \textbf{\textit{Stage 2:}} Fix $t$ and vary $s \sim \mathrm{U}$ (\textit{uniform distribution})
\For{$n = 0$ to $N-1$}
    \State Use weights $\mathbf{W}_2(\mathcal{F}_n) \gets \mathbf{W}_1(\mathcal{F}_n)$, $\mathbf{W}_2(\mathcal{U}_n) \gets \mathbf{W}_1(\mathcal{U}_n)$ 
    \For{$j = 0$ to $d_2-1$}
        \State Update all weights $\mathbf{W}_2(\mathcal{F}_n) \gets \mathbf{W}_2(\mathcal{F}_{n})$, $\mathbf{W}_2(\mathcal{U}_n) \gets \mathbf{W}_2(\mathcal{U}_{n})$
    \EndFor
\EndFor
\State Derive weights $\mathbf{W}^*(\mathcal{F}) \gets \mathbf{W}_2(\mathcal{F}_{N-1})$, $\mathbf{W}^*(\mathcal{U}_{N-1}) \gets \mathbf{W}_2(\mathcal{U}_{N-1})$
\State \textbf{\textit{Stage 3:}} Fix $t$ and vary $s \sim \mathrm{U}$ (\textit{uniform distribution})
\For{$n = 0$ to $N-2$}
    \State Use weights $\mathbf{W}_3(\mathcal{F}_n) \gets \mathbf{W}^*(\mathcal{F})$, $\mathbf{W}_3(\mathcal{U}_n) \gets \mathbf{W}_2(\mathcal{U}_n)$
    \For{$j = 0$ to $d_3-1$}
        \State Only update weights $\mathbf{W}_3(\mathcal{U}_n) \gets \mathbf{W}_3(\mathcal{U}_{n})$
    \EndFor
    \State Derive weights $\mathbf{W}^*(\mathcal{U}_n) \gets \mathbf{W}_3(\mathcal{U}_n)$
\EndFor
\Statex \hspace{-2em} \textbf{Output:} Target weights $\mathbf{W}^*(\mathcal{K}_n) \gets [\mathbf{W}^*(\mathcal{F});\{\mathbf{W}^*(\mathcal{U}_n)\}_{n=0}^{N-1}]$
\end{algorithmic}
\end{algorithm}

\section{Experiments}
\subsection{Experimental Setup}
\subsubsection{Datasets}
In our experiments, we train all models on the Adobe240 dataset~\cite{su2017deep} which contains 133 video sequences, where 100 for training, 16 for validation, and 17 for testing~\cite{chen2022videoinr}. We evaluate the performance of our models on both synthetic Adobe240~\cite{su2017deep} and GoPro~\cite{nah2017deep} datasets and real-world datasets including BS-ERGB \cite{tulyakov2022time} and ALPIX-VSR~\cite{lu2023learning}. For synthetic events, we use the event simulation method Vid2E \cite{gehrig2020video} to generate the event data. Given a spatial scale $s$ and a temporal scale $t$, we select consecutive $(t+1)$ frames as a clip. The first and last frames for $s:1$ bicubic downsampling are used as the LR and LFR inputs fed into the network to generate the HR and HFR frame sequences, respectively, and the original $(t+1)$ frames serve as the ground truth (GT). Following~\cite{kai2024evtexture}, the event voxels are bicubic-downsampled as the LR event inputs, thus ensuring the spatial size is aligned with the LR frames.

\begin{table}[t]
    \centering
    \caption{The detailed implementation settings of the cross-derivative training strategy (Algorithm~\ref{alg:1}) in our \textbf{EvEnhancerPlus}.}
    \centering
    \begin{tabular}{c|ccc}
    \hline
    Stages
    & Iterations
    & Temporal Scale
    & Spatial Scale
    \\
    \hline
    1
    & 450K
    & Fix $t=8$
    & Fix $s=4$
    \\
    2
    & 150K
    & Fix $t=8$
    & Vary $s \sim \mathrm{U}[1,4]$
    \\
    3
    & 150K
    & Fix $t=8$
    & Vary $s \sim \mathrm{U}[1,4]$
    \\
    \hline
    \end{tabular}
    \label{tab:Training_parameter_settings}
\end{table}

\newcolumntype{P}[1]{>{\centering\arraybackslash}p{#1}}

\begin{table*}
  \caption{Quantitative comparisons for in-distribution (\textit{In-Dist.}) spatiotemporal upsampling scales (\(t=8, s=4\)). \textbf{Bold} and \underline{underline} indicate the best and the second-best performance, respectively.}
  \centering
\begin{tabular}{cc|c|cccc|cccc|c}
    \hline
        \multirow{3}{*}{VFI Method}&
        \multirow{3}{*}{VSR Method}&
        \multirow{3}{*}{Events}&
        \multicolumn{4}{c|}{GoPro~\cite{nah2017deep}} &
        \multicolumn{4}{c|}{Adobe240~\cite{su2017deep}} &
        \multirow{3}{*}{Params (M)}\\
        &
        &
        &
        \multicolumn{2}{c}{-\emph{Center}} &
        \multicolumn{2}{c|}{-\emph{Average}} &
        \multicolumn{2}{c}{-\emph{Center}} &
        \multicolumn{2}{c|}{-\emph{Average}} &
        \\
        &
        &
        &
        PSNR &
        SSIM &
        PSNR &
        SSIM &
        PSNR &
        SSIM &
        PSNR &
        SSIM &
        \\
    \hline
        TimeLens~\cite{tulyakov2022time} &
        EGVSR~\cite{lu2023learning} &
        \checkmark &
        28.41 & 0.8307 &
        27.42 & 0.8077 &
        26.64 & 0.7675 &
        25.13 & 0.7298 &
        72.20+2.45 \\
        TimeLens~\cite{tulyakov2022time} &
        EvTexture~\cite{kai2024evtexture} &
        \checkmark &
        30.50 & 0.8784 &
        28.51 & 0.8478 &
        28.80 & 0.8337 &
        26.12 & 0.7789 &
        72.20+8.90 \\
        REFID~\cite{sun2023event} &
        EGVSR~\cite{lu2023learning} &
        \checkmark &
        28.74 & 0.8364 &
        27.97 & 0.8171 &
        27.28 & 0.7844 &
        26.55 & 0.7637 &
        15.91+2.45 \\
        REFID~\cite{sun2023event} &
        EvTexture~\cite{kai2024evtexture} &
        \checkmark &
        30.86 & 0.8784 &
        29.14 & 0.8489 &
        29.42 & 0.8424 &
        27.69 & 0.8059 &
        15.91+8.90 \\
        CBMNet-L~\cite{kim2023event} &
        EGVSR~\cite{lu2023learning} &
        \checkmark &
        29.08 & 0.8491 &
        29.39 & 0.8553 &
        27.97 & 0.8055 &
        27.66 & 0.7994 &
        22.23+2.45 \\
        CBMNet-L~\cite{kim2023event} &
        EvTexture~\cite{kai2024evtexture} &
        \checkmark &
        33.09 & 0.9227 &
        32.42 & 0.9171 &
        32.10 & 0.9074 &
        31.39 & 0.9014 &
        22.23+8.90 \\
    \hline
        \multicolumn{2}{c|}
        {Zooming Slow-Mo~\cite{huang2024scale}} &
        &
        30.69 & 0.8847 &
        - & - &
        30.26 & 0.8821 &
        - & - &
        11.10 \\
        \multicolumn{2}{c|}{TMNet~\cite{xu2021temporal}} &
        &
        30.14 & 0.8692 &
        28.83 & 0.8514 &
        29.41 & 0.8524 &
        28.30 & 0.8354 &
        12.26 \\
        \multicolumn{2}{c|}
        {VideoINR-\emph{fixed}~\cite{chen2022videoinr}} &
        &
        30.73 & 0.8850 &
        - & - &
        30.21 & 0.8805 &
        - & - &
        11.31 \\
        \multicolumn{2}{c|}
        {SAFA~\cite{huang2024scale}} &
        &
        31.28 & 0.8894 &
        30.22 & 0.8761 &
        30.97 & 0.8878 &
        30.13 & 0.8782 &
        4.94 \\
        \multicolumn{2}{c|}
        {EvSTVSR~\cite{yan2025evstvsr}} &
        \checkmark &
        32.50 & \textbf{0.9340} &
        32.23 & \underline{0.9320} &
        31.79 & \underline{0.9200} &
        31.61 & \underline{0.9194} &
        - \\
        \hline
        \multicolumn{2}{c|}{VideoINR~\cite{chen2022videoinr}} &
        &
        30.26 & 0.8792 &
        29.41 & 0.8669 &
        29.92 & 0.8746 &
        29.27 & 0.8651 &
        11.31 \\
        \multicolumn{2}{c|}{MoTIF~\cite{chen2023motif}} &
        &
        31.04 & 0.8877 &
        30.04 & 0.8773 &
        30.63 & 0.8839 &
        29.82 & 0.8750 &
        12.55 \\
        \multicolumn{2}{c|}{BF-STVSR~\cite{kim2025bf}} &
        &
        31.17 & 0.8898 &
        30.22 & 0.8802 &
        30.83 & 0.8880 &
        30.12 & 0.8808 &
        13.47 \\
        \multicolumn{2}{c|}{HR-INR~\cite{lu2024hr}} &
        \checkmark &
        31.97 & 0.9298 &
        32.13 & \textbf{0.9371} &
        31.26 & \textbf{0.9246} &
        31.11 & \textbf{0.9216} &
        8.27 \\
    \hline
        \multicolumn{2}{c|}{\textbf{EvEnhancer} \textbf{(Ours)}} &
        \checkmark &
        \underline{33.52} & 0.9295 &
        \underline{33.30} & 0.9279 &
        \textbf{32.43} & 0.9129 &
        \underline{32.18} & 0.9116 &
        6.55 \\
        \rowcolor{gray!20}
        \multicolumn{2}{c|}{\textbf{EvEnhancerPlus} \textbf{(Ours)}} &
        \checkmark &
        \textbf{33.57} & \underline{0.9303} &
        \textbf{33.39} & 0.9291 &
        \underline{32.42} & 0.9127 &
        \textbf{32.21} & 0.9120 &
        6.91 \\
    \hline
  \end{tabular}
  \label{tab:Quantitative_comparisons_T8S4}
\end{table*}
\begin{table*}
  \caption{Quantitative comparisons with C-STVSR methods for out-of-distribution (\textit{OOD}) spatiotemporal upsampling scales on the GoPro dataset \cite{nah2017deep}. 
  \textbf{Bold} and \underline{underline} indicate the best and the second-best performance, respectively.
  }
  \centering
  \resizebox{\textwidth}{!}{
  \begin{tabular}{cc||cccccccc|cc>{\columncolor{gray!20}}c>{\columncolor{gray!20}}c}
    \hline
        Temporal & 
        Spatial & 
        \multicolumn{2}{c}{VideoINR~\cite{chen2022videoinr}} &
        \multicolumn{2}{c}{MoTIF~\cite{chen2023motif}} &
        \multicolumn{2}{c}{BF-STVSR~\cite{kim2025bf}} &
        \multicolumn{2}{c|}{HR-INR~\cite{lu2024hr}} &
        \multicolumn{2}{c}{\textbf{EvEnhancer}} &
        \multicolumn{2}{>{\columncolor{gray!20}}c}{\textbf{EvEnhancerPlus}} \\
        Scale & 
        Scale & 
        PSNR & SSIM &
        PSNR & SSIM &
        PSNR & SSIM &
        PSNR & SSIM &
        PSNR & SSIM &
        PSNR & SSIM
        \\
    \hline
        $t=6$ & 
        $s=4$ &
        30.78 & 0.8954 &
        31.56 & 0.9064 &
        31.70 & 0.9083 &
        - & - &
        \underline{33.41} & \underline{0.9300} &
        \textbf{33.46} & \textbf{0.9304} 
        \\
        $t=6$ & 
        $s=6$ &
        25.56 & 0.7671 &
        29.36 & 0.8505 &
        29.45 & 0.8520 &
        - & - &
        \underline{30.12} & \textbf{0.8675} &    
        \textbf{30.15} & \underline{0.8672}
        \\
        $t=6$ & 
        $s=12$ &
        24.02 & 0.6900 &
        \textbf{25.81} & \textbf{0.7330} &
        25.80 & 0.7295 &
        - & - &
        25.50 & \underline{0.7323} &
        \underline{25.57} & \underline{0.7323}
        \\
    \hline
        $t=12$ & 
        $s=4$ &
        27.32 & 0.8141 &
        27.77 & 0.8230 &
        28.07 & 0.8287 &
        28.87 & 0.8854 &
        \underline{32.07} & \underline{0.9116} &
        \textbf{32.19} & \textbf{0.9137}
        \\
        $t=12$ & 
        $s=6$ &
        24.68 & 0.7358 &
        26.78 & 0.7908 &
        27.07 & 0.7963 &
        27.14 & 0.8173 &
        \underline{29.54} & \underline{0.8579} &
        \textbf{29.63} & \textbf{0.8591}
        \\
        $t=12$ & 
        $s=12$ &
        23.70 & 0.6830 &
        24.72 & 0.7108 &
        24.88 & 0.7104 &
        - & - &
        \underline{25.45} & \underline{0.7345} &        
        \textbf{25.53} & \textbf{0.7351}
        \\
    \hline
        $t=16$ & 
        $s=4$ &
        25.81 & 0.7739 &
        25.98 & 0.7758 &
        26.39 & 0.7840 &
        27.29 & 0.8556 &
        \underline{30.93} & \underline{0.8918} &   
        \textbf{31.11} & \textbf{0.8954}
        \\
        $t=16$ & 
        $s=6$ &
        23.86 & 0.7123 &
        25.34 & 0.7527 &
        25.81 & 0.7619 &
        26.09 & 0.7954 &
        \underline{28.86} & \underline{0.8434} &
        \textbf{29.00} & \textbf{0.8460}
        \\
        $t=16$ & 
        $s=12$ &
        22.88 & 0.6659 &
        23.88 & 0.6923 &
        24.22 & 0.9655 &
        - & - &
        \underline{25.25} & \underline{0.7310} &
        \textbf{25.36} & \textbf{0.7320} 
        \\
    \hline
        $t=6$ & 
        $s=1$ &
        32.34 & 0.9545 &
        34.77 & 0.9696 &
        33.96 & 0.9272 &
        \underline{38.53} & \textbf{0.9735} &
        \textbf{38.80} & 0.9714 &
        \underline{38.53} & \underline{0.9728}
        \\
    \hline
  \end{tabular}}
  \label{tab:Quantitative_comparisons_TxSx}
\end{table*}

\subsubsection{Implementation Details}
In both \textbf{EvEnhancer} and \textbf{EvEnhancerPlus},  we set the number of event segments $M=7$. In EASM, each convolutional block comprises a $3\times 3$ kernel with 64 channels activated by LeakyReLU. There are two convolutional blocks in all residual blocks. In LIVT, the local grid has the size of $T^{G} \times H^{G} \times W^{G}=3 \times 3 \times 3$ and all 3D convolutional layers have the kernel size of $3\times 3\times 3$. The MLP with a hidden dimension of 256 uses GELU activation. For \textbf{EvEnhancer}, 
Referring to VideoINR~\cite{chen2022videoinr}, the training for our model comprises two stages. In the first stage, we set $s=4$ and $t=8$, training for 450K iterations on Adobe240~\cite{su2017deep}. In the second stage, we uniformly sample $s$ in $[1, 4]$ and fine-tune for another 150K iterations on the same dataset.

In \textbf{EvEnhancerPlus}, there are two upsamplers with different configurations: one is a lightweight LIVT $\mathcal{U}_0$ with convolutional layers of 16 channels and the other is the original LIVT $\mathcal{U}_1$ in \textbf{EvEnhancer} (64 channels). In this case, there is an adjustable threshold $\xi_1 \in [0,1]$ in CSM. The detailed settings of the cross-derivative training strategy are presented in Table~\ref{tab:Training_parameter_settings}. 

We use the Adam optimizer \cite{kingma2014adam} with \(\beta_{1}=0.9\) and \(\beta_{2}=0.999\), and apply cosine annealing to decay the learning rate from \(1 \times 10^{-4}\) to \(1 \times 10^{-7}\) over 150K iterations. For data augmentation, we randomly crop \(32 \times 32\) spatial regions from downsampled images and event voxel grids with random rotations and horizontal flips. The model is optimized with Charbonnier loss \cite{lai2017deep} using \(\epsilon^2=1 \times 10^{-6}\) with a batch size of 4. The entire training process runs on NVIDIA GeForce RTX 3090 GPUs. To measure our models, we calculate the PSNR (dB) and SSIM on the Y channel.

\begin{figure*}[t]
\centering
\includegraphics[width=\textwidth]{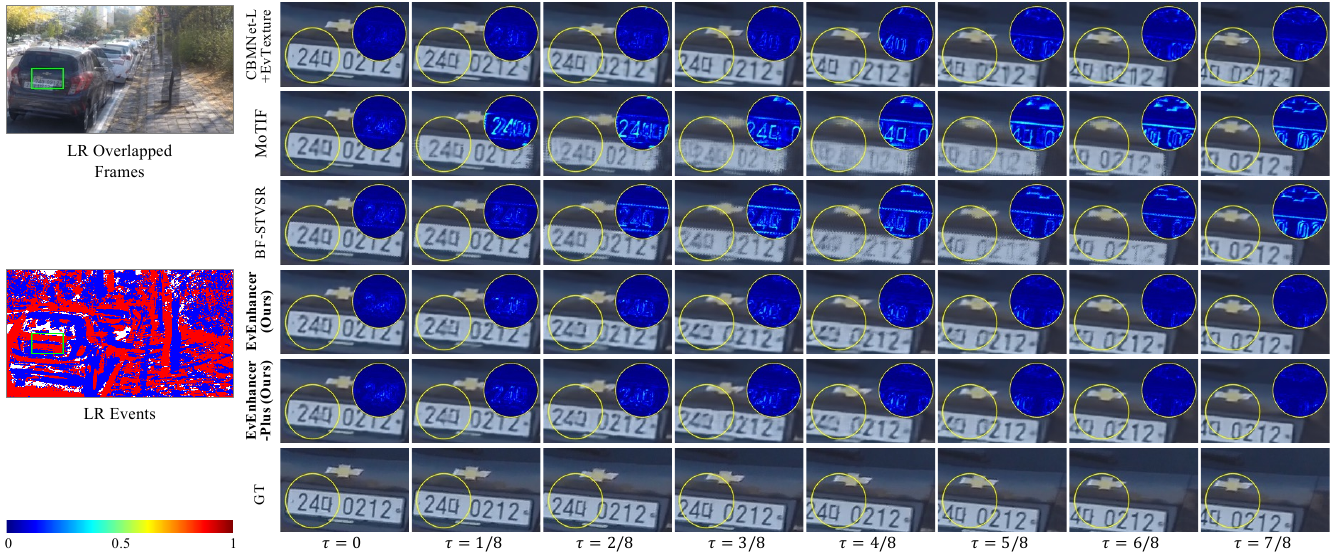}
\caption{Qualitative comparison for \textit{In-Dist.} scale ($t=8$, $s=4$) on the GoPro dataset \cite{nah2017deep}. Residual intensity maps corresponding to the same regions in each frame as those in the GT frames.}
\label{fig:Qualitative_comparisons_a}
\end{figure*}

\subsection{Comparison with State-of-the-Art Methods}
We comprehensively consider frame-based and event-based one-stage or two-stage state-of-the-art (SOTA) methods for comparison: 
1) two-stage event-based STVSR methods consisting of event-based VFI (TimeLens \cite{tulyakov2021time}, REFID \cite{sun2023event}, CBMNet-L \cite{kim2023event}) and event-based VSR (EGVSR \cite{lu2023learning}, EvTexture \cite{kai2024evtexture});   
2) one-stage frame-based F-STVSR methods (Zooming Slow-Mo~\cite{xiang2020zooming}, TMNet \cite{xu2021temporal}, VideoINR-\emph{fixed} \cite{chen2022videoinr}, SAFA \cite{huang2024scale}), and event-based ones (EvSTVSR \cite{yan2025evstvsr});
3) one-stage frame-based C-STVSR methods (VideoINR \cite{chen2022videoinr}, MoTIF \cite{chen2023motif}, BF-STVSR \cite{kim2025bf}), and event-based ones (HR-INR \cite{lu2024hr}).
For fair comparison as \cite{chen2022videoinr}, all VFI and STVSR models are trained on the same Adobe240 dataset~\cite{su2017deep}, unless otherwise specified. Due to the unavailability of source codes for certain methods, we adopt the officially released pre-trained model of TimeLens~\cite{tulyakov2021time}, while the results of EvSTVSR~\cite{yan2025evstvsr} and~HR-INR \cite{lu2024hr} are cited from their original papers.

\subsubsection{Quantitative Comparison}
Table~\ref{tab:Quantitative_comparisons_T8S4} presents the results for the F-STVSR task, where the temporal scale $t=8$ and the spatial scale $s=4$ fall within the training distribution (\textit{In-Dist.}). ``-\emph{Center}'' represents the average performance of the reconstructed left, right, and center frames. ``-\emph{Average}'' represents the average performance of all reconstructed frames. It can be observed that both \textbf{EvEnhancer} and \textbf{EvEnhancerPlus} significantly outperform most existing methods in all settings with much fewer model parameters, which achieve more than 0.5 dB PSNR gains. By using CSM in \textbf{EvEnhancerPlus} to adaptively route pixels to different upsamplers, it demonstrates consistent improvements in most cases.

Then, we also evaluate our models on the C-STVSR task, where spatial and temporal scales are out of the training distribution (\textit{OOD}). The results are averaged over all reconstructed frames and illustrated in~Table~\ref{tab:Quantitative_comparisons_TxSx}. Benefiting from the cross-scale attention in LIVT, both our models exceed almost all existing methods at \textit{OOD} scales, especially at the challenging large temporal scale, verifying superior effectiveness and generalization ability.

\subsubsection{Model Complexity}
To validate the model efficiency of the proposed methods, in addition to the parameter comparison in Table~\ref{tab:Quantitative_comparisons_T8S4}, we compare the TFLOPs with existing C-STVSR methods~\cite{chen2022videoinr,chen2023motif,kim2025bf} on the GoPro dataset~\cite{nah2017deep} in Table~\ref{tab:flops}. We calculate on the upsampling of a clip with a resolution of 180 \(\times\) 320 by a spatial upsampling scale \(s=4\) at different temporal upsampling scales \(t\). As shown, VideoINR~\cite{chen2022videoinr} shows the fewest TFLOPs but more parameters and the worst performance (see Tables~\ref{tab:Quantitative_comparisons_T8S4} and \ref{tab:Quantitative_comparisons_TxSx}). MoTIF~\cite{chen2023motif} and BF-STVSR~\cite{kim2025bf} are more computationally efficient than ours at low temporal scales ($t \leq 4$). However, as the temporal scale increases, their computational costs increase and exceed both our models. We can also see that \textbf{EvEnhancerPlus} is obviously more computationally efficient than \textbf{EvEnhancer}, saving about 12.8\% TFLOPs. This can be attributed to the adaptive pathways design in CSM based on the difficulty in pixel reconstruction. The overall comparisons validate that our methods can achieve an optimal trade-off between effectiveness, generalizability, and efficiency.

\begin{table}[t]
    \centering
    \caption{TFLOPs comparisons of C-STVSR models on the GoPro dataset~\cite{nah2017deep} at different temporal upsampling  scales \(t\) (fixed \(s=4\)). $^\dagger$ indicates the minimum repeated iterations during inference on a single NVIDIA GeForce RTX 3090 GPU due to their large computations.}
    \resizebox{\columnwidth}{!}{\begin{tabular}{c|cccccc}
    \hline
    Method & $t=2$ & $t=4$ & $t=6$ & $t=8$ & $t=12$ & $t=16$ \\
    \hline
    VideoINR~\cite{chen2022videoinr} & 2.011 & 2.395 & 2.779 & 3.163 & 3.932 & 4.700 \\
    MoTIF~\cite{chen2023motif} & 2.043 & 4.004$^\dagger$ & 5.965$^\dagger$ & 6.129$^\dagger$ & 10.05$^\dagger$ & 12.18$^\dagger$ \\
    BF-STVSR~\cite{kim2025bf} & 1.896 & 3.756$^\dagger$ & 5.616$^\dagger$ & 5.686$^\dagger$ & 9.407$^\dagger$ & 11.34$^\dagger$ \\
    \hline
    \textbf{EvEnhancer (Ours)} & 3.398 & 4.647 & 5.885 & 7.129 & 9.617 & 12.10 \\
    \rowcolor{gray!20}
    \textbf{EvEnhancerPlus (Ours)} & 2.967 & 4.003 & 5.004 & 6.125 & 8.339 & 10.55 \\
    \hline
    \end{tabular}}
    \label{tab:flops}
\end{table}

\begin{table*}[t]
    \centering
    \caption{Quantitative comparisons for different temporal upsampling scales \(t\) (fixed \(s=4\)) on the BS-ERGB dataset~\cite{tulyakov2022time}. \textbf{Bold} and \underline{underline} indicate the best and the second-best performance, respectively.}
    \centering
    \begin{tabular}{cc|c|cccccc}
    \hline
        \multirow{2}{*}{VFI Method}&
        \multirow{2}{*}{VSR Method} &
        \multirow{2}{*}{Events} &
        \multicolumn{2}{c}{\(t=4\)} &
        \multicolumn{2}{c}{\(t=6\)} &
        \multicolumn{2}{c}{\(t=8\)} \\
        &
        &
        &
        PSNR & SSIM &
        PSNR & SSIM &
        PSNR & SSIM\\
    \hline
        TimeLens~\cite{tulyakov2022time} & EGVSR~\cite{lu2023learning} &
        \checkmark &
        22.98 & 0.6762 &
        22.01 & 0.6592 &
        21.33 & 0.6467 \\
        TimeLens~\cite{tulyakov2022time} & EvTexture~\cite{kai2024evtexture} &
        \checkmark &
        23.58 & 0.7088 &
        22.48 & 0.6887 &
        21.74 & 0.6746\\
        REFID~\cite{sun2023event} & EGVSR~\cite{lu2023learning} &
        \checkmark &
        - & - &
        - & - &
        22.76 & 0.6478 \\
        REFID~\cite{sun2023event} & EvTexture~\cite{kai2024evtexture} &
        \checkmark &
        - & - &
        - & - &
        22.77 & 0.6527 \\
        CBMNet-L~\cite{kim2023event} & EGVSR~\cite{lu2023learning} &
        \checkmark &
        23.60 & 0.6852 &
        22.76 & 0.6716 &
        22.15 & 0.6605 \\
        CBMNet-L~\cite{kim2023event} & EvTexture~\cite{kai2024evtexture} &
        \checkmark &
        24.13 & 0.7141 &
        23.14 & 0.6972 &
        22.46 & 0.6846\\
    \hline
        \multicolumn{2}{c|}{SAFA~\cite{huang2024scale}} &
        &
        24.32 & 0.7314 &
        23.23 & 0.7120 &
        22.41 & 0.6954\\
        \multicolumn{2}{c|}{VideoINR~\cite{chen2022videoinr}} &
        &
        24.06 & 0.7290 &
        23.00 & 0.7098 &
        22.23 & 0.6945 \\
        \multicolumn{2}{c|}{MoTIF~\cite{chen2023motif}} &
        &
        24.21 & 0.7296 &
        23.15 & 0.7099 &
        22.37 & 0.6939 \\
        \multicolumn{2}{c|}{BF-STVSR~\cite{kim2025bf}} &
        &
        24.05 & 0.7259 &
        22.99 & 0.7062 &
        22.19 & 0.6896 \\
    \hline
        \multicolumn{2}{c|}{\textbf{EvEnhancer (Ours)}}&
        \checkmark &
        \textbf{25.44} & \textbf{0.7338} &
        \underline{24.70} & \underline{0.7215} &
        \underline{24.29} & \underline{0.7138}\\
        \rowcolor{gray!20}
        \multicolumn{2}{c|}{\textbf{EvEnhancerPlus (Ours)}} &
        \checkmark &
        \underline{25.40} & \underline{0.7326} &
        \textbf{24.71} & \textbf{0.7216} &
        \textbf{24.35} & \textbf{0.7146} \\
    \hline
  \end{tabular}
  \label{tab:bsergb}
\end{table*}
\begin{table*}[ht]
    \centering
    \caption{Quantitative comparisons with methods that enable arbitrary VSR for scale $s=2$ on the ALPIX-VSR dataset~\cite{lu2023learning}. \textbf{Bold} and \underline{underline} indicate the best and the second-best performance, respectively.}
    \centering
    \begin{tabular}{c|cccc|c>{\columncolor{gray!20}}c}
    \hline
    Method
    & VideoINR~\cite{chen2022videoinr}
    & MoTIF~\cite{chen2023motif}
    & BF-STVSR~\cite{kim2025bf}
    & EGVSR~\cite{lu2023learning}
    & \textbf{EvEnhancer}
    & \textbf{EvEnhancerPlus}
    \\
    \hline
    PSNR
    & 32.53
    & 38.61
    & 39.27
    & 39.31
    & \underline{40.84}
    & \textbf{42.01}
    \\
    SSIM
    & 0.9383
    & 0.9636
    & 0.9657
    & 0.9635
    & \underline{0.9786}
    & \textbf{0.9802}
    \\
    \hline
    \end{tabular}
    \label{tab:alpix}
\end{table*}

\subsubsection{Qualitative Comparison}

\label{sec:Qualitative_Comparison}
Fig. \ref{fig:Qualitative_comparisons_a} presents qualitative results on the GoPro dataset~\cite{nah2017deep} at \textit{In-Dist.} scales (temporal scale $t=8$ and spatial scale $s=4$), along with residual intensity maps between the reconstructed frames and their corresponding GT frames. It can be observed that frame-based SOTA methods MoTIF~\cite{chen2023motif} and BF-STVSR~\cite{kim2025bf} suffer from noticeable quality degradation when reconstructing intermediate frames. Though event-based SOTA methods CBMNet-L~\cite{kim2023event}+EvTexture~\cite{kai2024evtexture} perform slightly better, their generated images still contain visually unpleasant artifacts. In contrast, our models can produce faithful HR frames with clearer textures. The resultant residual intensity maps also reveal that our models are better at tackling non-linear motions and recovering more details.

\begin{figure*}[t]
\centering
\includegraphics[width=\textwidth]{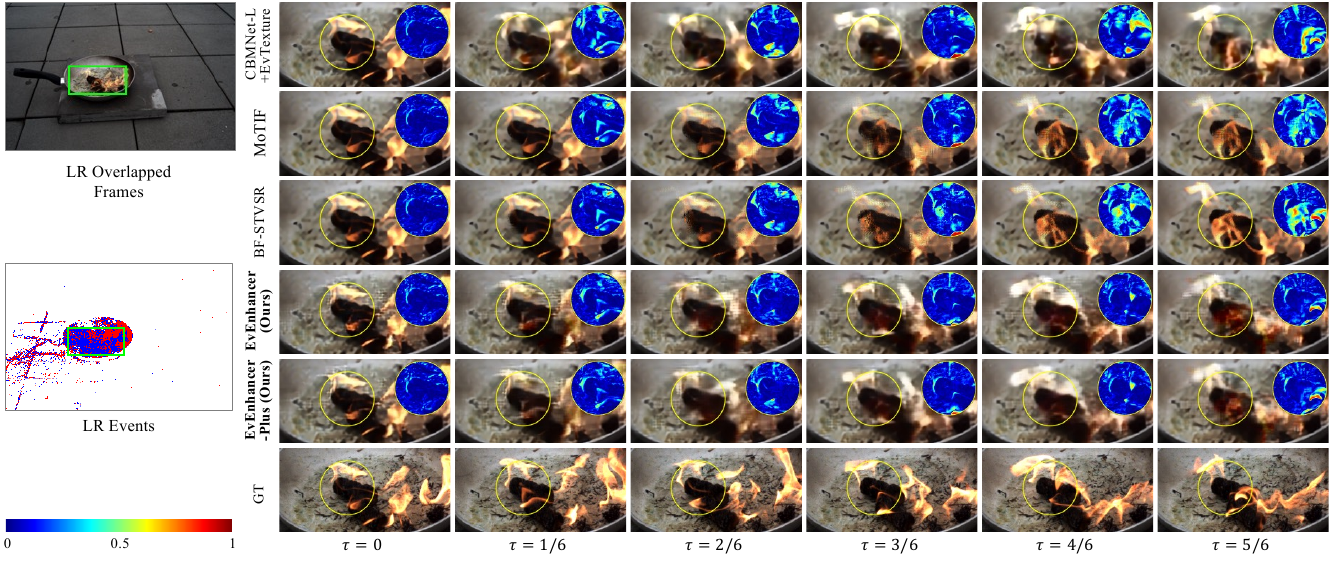}
\caption{Qualitative comparison for \textit{OOD} scale ($t=6$, $s=4$) on the BS-ERGB dataset \cite{tulyakov2022time}. Residual intensity maps corresponding to the same regions
in each frame as those in the GT frames.}
\label{fig:Qualitative_comparisons_b}
\end{figure*}

\subsubsection{Experiments on Real-World Datasets}
We validate the effectiveness of our method on the real-world BS-ERGB dataset~\cite{tulyakov2022time} under the setting of fixed spatial scale \(s=4\) but varying temporal scales \(t\). The comparisons in Table~\ref{tab:bsergb} demonstrate that our method yields the best quantitative performance under all conditions. Compared to the cascaded event-based VFI and VSR methods, our method simultaneously learns temporal interpolation and spatial super-resolution, thus achieving outstanding performance. Due to the high temporal resolution of events, equipped by EASM, our method enables holistic motion trajectory modeling, thus outperforming frame-based methods at both \textit{In-Dist.} and \textit{OOD} scales. 

We further investigate the performance of our method on real-world LR events by conducting VSR evaluations with a spatial scale $s=2$ on the real-world ALPIX-VSR dataset~\cite{lu2023learning}. The raw events have a resolution that is half of the corresponding images. We select the central $1024 \times 1024$ region of each image as GT, while the inputs consist of the corresponding raw $512 \times 512$ LR events and $2\times$ downsampled images. As shown in Table~\ref{tab:alpix}, our two models significantly surpass all existing methods without additional fine-tuning, validating the preferable robustness.

\begin{figure*}[t]
\centering
\includegraphics[width=\textwidth]{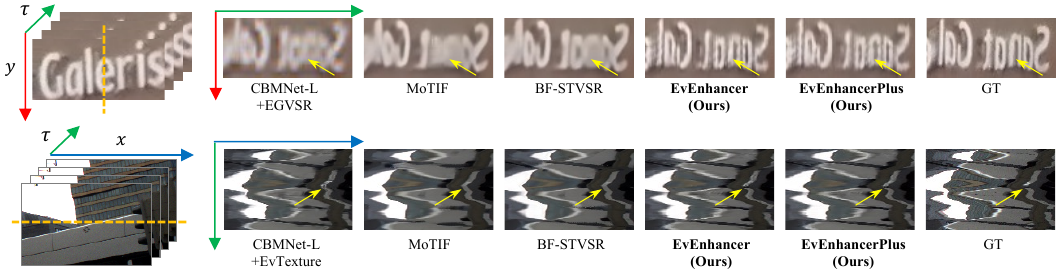}
\caption{Comparison of temporal profiles across different datasets and scales. The upper shows the temporal profile on the GoPro dataset~\cite{nah2017deep} ($t=12$, $s=6$), where a column (indicated by orange dotted lines) is selected to observe changes over time. The lower presents the temporal profile on the BS-ERGB dataset~\cite{tulyakov2022time} ($t=4$, $s=4$), where a row (indicated by orange dotted lines) is selected to observe temporal changes.}
\label{fig:temporal_profile}
\end{figure*}

Fig. \ref{fig:Qualitative_comparisons_b} illustrates reconstruction results on the BS-ERGB dataset~\cite{tulyakov2022time} at \textit{OOD} scales (temporal scale $t=6$ and spatial scale $s=4$). In this highly challenging flame combustion scenario, all compared SOTA methods fail to accurately reconstruct the evolution of the flame, whereas our \textbf{EvEnhancer} and \textbf{EvEnhancerPlus} are still able to effectively capture clear and dynamic flame-burning details, exhibiting smaller deviations from the GT frames.

\subsubsection{Temporal Consistency}
In Fig.~\ref{fig:temporal_profile}, we present temporal profile visualizations across different datasets and reconstruction scales to evaluate temporal consistency. As shown, MoTIF~\cite{chen2023motif} and BF-STVSR~\cite{kim2025bf} suffer from severe flickering artifacts, noise, and blurs across frames, indicating their poor temporal coherence. Although the event-based VFI + VSR pipelines (CBMNet-L~\cite{kim2023event} + EGVSR~\cite{lu2023learning} and CBMNet-L~\cite{kim2023event} + EvTexture~\cite{kai2024evtexture}) can better handle motion, they still struggle with temporal inconsistencies and accumulated errors. In contrast, our \textbf{EvEnhancer} shows more pleasant temporal profiles but contains discontinuities and artifacts. Moreover, the \textbf{EvEnhancerPlus} delivers the most temporally coherent results with smooth transitions, maintaining the best consistency.

\subsection{Ablation Studies}
In this subsection, we conduct comprehensive ablation studies to validate the effectiveness of our methodology design. Firstly, we take the original \textbf{EvEnhancer} as the baseline and investigate the individual contribution of each module in the network architecture (Sec.~\ref{sec:ablation_network}). 
Then, we study the effect of the proposed CSM and cross-derivative training strategy in \textbf{EvEnhancerPlus} (Sec.~\ref{sec:ablation_improvement}). All ablation experiments are conducted with models trained on the Adobe240 dataset~\cite{su2017deep}. For evaluation, we define a temporal scale of $t=8$ and a spatial scale of $s=4$ on the GoPro dataset~\cite{nah2017deep} to represent in-distribution (\textit{In-Dist.}) conditions, whereas $t=12$ and $s=6$ are used to represent out-of-distribution (\textit{OOD}) conditions, unless otherwise specified.

\subsubsection{Network Architecture}
\label{sec:ablation_network}
\textbf{Event-Adapted Synthesis Module (EASM).} 
Table \ref{tab:Ab_EASM} demonstrates the effects of EASM design,  which involves the event-modulated alignment (EMA), bidirectional recurrent compensation (BRC), and bidirectional feature fusion (BF, Eq.~(\ref{eq:5-1})). The model with only EMA achieves the worst performance. When we introduce BRC in the model, we can see the improvement in performance. Besides, with the integration of EMA, BRC, and BF, the model achieves the best at both \textit{In-Dist.} and \textit{OOD} conditions. We can also observe that the model with only a single forward alignment can suffer from performance degradation. 

Next, in EMA, we introduce multi-scale event and image features for alignment. Table \ref{tab:Ab_EMA} shows that 1$\times$ scale alignment plays a more critical role. However, the multi-scale manner (\(1\times\), \(\frac{1}{2}\times\), \(\frac{1}{4}\times\)) captures richer motion cues compared to the single-scale setting, leading to improved performance. Table \ref{tab:Ab_BRC} presents the further ablations of BRC, which involve bidirectional (forward and backward) recurrence in BRC. 

We also investigate the influence of channel attention mechanisms in BRC. As we can see in Table \ref{tab:Ab_BRC}, the model using only the forward or backward compensation shows the worst performance. When we incorporate attention in each direction, the performance increases. Moreover, by implementing bidirectional compensation with attention both forward and backward, the model performs the best.

\begin{table}[t]
    \centering
    \caption{Ablation studies on the designs of event-adapted synthesis module (EASM). ``fwd.'' and ``bwd.'' denote the forward and backward directions, respectively.}
    \centering
    \begin{tabular}{ccc||cc|cc}
    \hline
    \multirow{2}{*}{EMA}
    & \multirow{2}{*}{BRC}
    & \multirow{2}{*}{BF}
    & \multicolumn{2}{c|}{\textit{In-Dist.}}
    & \multicolumn{2}{c}{\textit{OOD}}\\
    &
    &
    & PSNR
    & SSIM
    & PSNR
    & SSIM
    \\
    \hline
    fwd. \& bwd.
    & 
    & 
    & 32.50 & 0.9166
    & 28.86 & 0.8407\\
    fwd. \& bwd.
    & \checkmark
    & 
    & 33.13 & 0.9252
    & 29.42 & 0.8539\\
    fwd.
    & \checkmark
    & \checkmark
    & 33.20 & 0.9262
    & 29.47 & 0.8556\\
    \rowcolor{gray!20}
    fwd. \& bwd.
    & \checkmark
    & \checkmark
    & \textbf{33.30} & \textbf{0.9279}
    & \textbf{29.54} & \textbf{0.8579}\\
    \hline
    \end{tabular}
    \label{tab:Ab_EASM}
\end{table}
\begin{table}[t]
    \centering
    \caption{Ablation studies on the event-modulated alignment (EMA).}
    \centering
    \begin{tabular}{c||cc|cc}
    \hline
     \multirow{2}{*}{Alignment}
    & \multicolumn{2}{c|}{\textit{In-Dist.}}
    & \multicolumn{2}{c}{\textit{OOD}}\\
    & PSNR & SSIM
    & PSNR & SSIM \\
    \hline
    1$\times$ Scale
    & 33.21 & 0.9266 
    & 29.46 & 0.8560 \\
    1/2$\times$ Scale
    & 32.98 & 0.9231 
    & 29.32 & 0.8520 \\
    1/4$\times$ Scale
    & 32.92 & 0.9223
    & 29.26 & 0.8502 \\
    \rowcolor{gray!20}
    Multi-Scale
    & \textbf{33.30} & \textbf{0.9279}
    & \textbf{29.54} & \textbf{0.8579}\\
    \hline
    \end{tabular}
    \label{tab:Ab_EMA}
\end{table}
\begin{table}[t]
    \centering
    \caption{Ablation studies on the bidirectional recurrent compensation (BRC).}
    \centering
    \begin{tabular}{cc||cc|cc}
    \hline
    \multirow{2}{*}{Recurrent}
    & Channel
    & \multicolumn{2}{c|}{\textit{In-Dist.}}
    & \multicolumn{2}{c}{\textit{OOD}}\\
    & Attention
    & PSNR
    & SSIM
    & PSNR
    & SSIM\\
    \hline
    fwd.
    & 
    & 32.57 & 0.9176 
    & 28.96 & 0.8429 \\
    bwd.
    & 
    & 32.55 & 0.9171 
    & 28.94 & 0.8424 \\
    fwd.
    & \checkmark
    & 32.79 & 0.9211 
    & 29.12 & 0.8467 \\
    bwd.
    & \checkmark
    & 32.79 & 0.9207 
    & 29.11 & 0.8467 \\
    \rowcolor{gray!20}
    fwd. \& bwd.
    & \checkmark
    & \textbf{33.30} & \textbf{0.9279}
    & \textbf{29.54} & \textbf{0.8579}\\
    \hline
    \end{tabular}
    \label{tab:Ab_BRC}
\end{table}

\begin{figure*}[t]
\centering
\includegraphics[width=\textwidth]{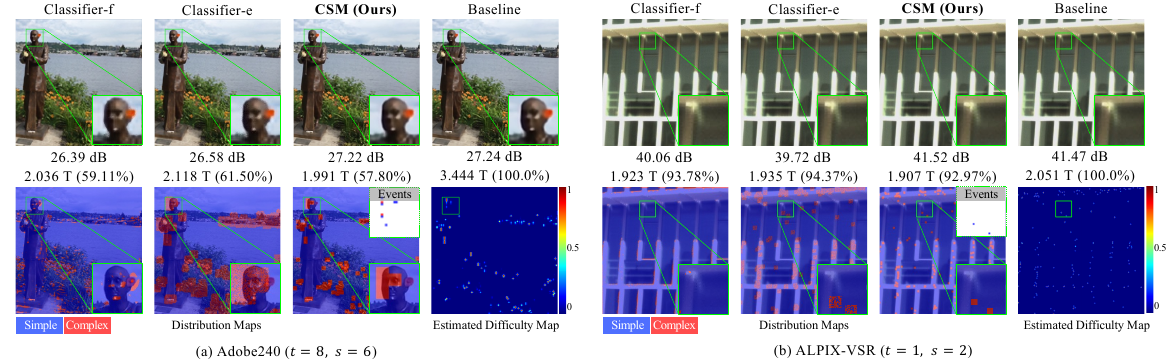}
\caption{Qualitative results of reconstruction performance (the top row), the distribution maps among different distributors, and the reconstruction difficulty map estimated by our CSM from events (the bottom row) on (a) Adobe240~\cite{su2017deep} (center frame visualization) and (b) ALPIX-VSR~\cite{lu2023learning}. Metrics: PSNR and FLOPs (ratio). Best zoom in for better visualization.}
\label{fig:distribution}
\end{figure*}

\begin{table}[t]
    \centering
    \caption{Ablation studies on the designs of local implicit video transformer (LIVT). ``Decpl.'' and ``Unifi.'' denote the decoupling and unification modes, respectively. ``LA": local attention, ``CE": concatenate cell decoding~\cite{chen2021learning}, ``PQ": concatenate the previous query,}
    \centering
    \resizebox{\columnwidth}{!}{\begin{tabular}{cccc||cc|cc}
    \hline
    INR
    & \multirow{2}{*}{LA}
    & \multirow{2}{*}{CE}
    & \multirow{2}{*}{PQ}
    & \multicolumn{2}{c|}{\textit{In-Dist.}}
    & \multicolumn{2}{c}{\textit{OOD}}\\
    Mode & 
    &
    & 
    & PSNR & SSIM
    & PSNR & SSIM\\
    \hline
    2D Decpl.
    & \checkmark
    & \checkmark
    & \checkmark
    & 33.26 & 0.9272
    & 29.26 & 0.8525\\
    3D Unifi.   
    &
    &
    &
    & 27.32 & 0.7975
    & 25.03 & 0.7032\\
    3D Unifi. 
    & \checkmark
    &
    &
    & 33.27 & 0.9276
    & 29.47 & 0.8565\\
    3D Unifi. 
    & \checkmark
    & \checkmark
    & 
    & 33.30 & \textbf{0.9282}
    & 29.49 & 0.8570\\
    \hline
    3D Unifi. 
    & \checkmark
    & \checkmark
    & \checkmark
    & 33.30 & 0.9279
    & \textbf{29.54} & \textbf{0.8579}\\
    \rowcolor{gray!20}
    3D Unifi. 
    & \checkmark
    & 
    & \checkmark
    & \textbf{33.31} & \textbf{0.9282}
    & \textbf{29.54} & 0.8578\\
    \hline
    \end{tabular}}
    \label{tab:Ab_LIVT}
\end{table}
\begin{table}[t]
    \centering
    \caption{Ablation studies on the attention mechanism and positional encoding in LIVT.}
    \centering
    \resizebox{\columnwidth}{!}{
    \begin{tabular}{cc||cc|cc}
    \hline
    Attention
    & Positional
    & \multicolumn{2}{c|}{\textit{In-Dist.}}
    & \multicolumn{2}{c}{\textit{OOD}}\\
    Mechanism
    & Encoding
    & PSNR
    & SSIM
    & PSNR
    & SSIM\\
    \hline
    Neighborhood~\cite{hassani2023neighborhood}
    & Cosine
    & 33.07 & 0.9248
    & 29.24 & 0.8492
    \\
    Cross-Scale
    & Learnable~\cite{mildenhall2021nerf}
    & 31.92 & 0.9086 
    & 28.37 & 0.8294 \\
    \rowcolor{gray!20}
    Cross-Scale
    & Cosine
    & \textbf{33.30} & \textbf{0.9279}
    & \textbf{29.54} & \textbf{0.8579}\\
    \hline
    \end{tabular}}
    \label{tab:Ab_Attention}
\end{table}

\textbf{Local Implicit Video Transformer (LIVT).} Table~\ref{tab:Ab_LIVT} investigates the effect of LIVT design. We first implement a model based on the 2D decoupling strategy from VideoINR~\cite{chen2022videoinr}, while incorporating local attention, cell decoding concatenation~\cite{chen2021learning}, and the previous query concatenation~\cite{chen2022videoinr}. This configuration already outperforms the baseline model without these components. When we replace the 2D decoupling with our unified video INR, the model achieves improvements particularly under \textit{OOD} conditions. The concatenation of cell decoding and the previous query yields greater performance gains at \textit{OOD} scales compared to \textit{In-Dist.} settings, demonstrating that retaining these components remains effective. Notably, the addition of local attention significantly enhances performance, demonstrating its importance in learning an accurate unified video INR. We further observe that removing the cell decoding concatenation from the original LIVT in \textbf{EvEnhancer} has a negligible impact under \textit{OOD} conditions, and even leads to performance improvements under \textit{In-Dist.} settings, thus we adopt this modification in the extended version.

In LIVT, the local attention is calculated from cross-scale features, \emph{i.e.}, the query is obtained from the large-scale features at the HR grid after trilinear upsampling, while the key and value are obtained from the small-scale at the local LR grids nearest to the query. Our cross-scale attention can be seen as a 3D cross-scale derivation of neighborhood attention~\cite{hassani2023neighborhood}. As illustrated in Table~\ref{tab:Ab_Attention}, it exhibits suboptimal performance if we use this neighborhood attention directly on the same feature scale. 

Besides, we encode and reshape the spatiotemporal relative coordinates $(\delta \tau, \delta x, \delta y)\in(-1,1)$ from each query point to all pixel points within its local grid via the cosine positional encoding~(Eq.~(\ref{eq:7})). Here, we investigate its impact by comparing it with another learnable positional encoding scheme~\cite{mildenhall2021nerf}. Table~\ref{tab:Ab_Attention} shows that the model with cosine positional encoding is superior to the learnable one.

\textbf{Hyperparameter Analysis.} Here, we investigate hyperparameters including the number of event segments \(M\) in EASM, the local grid size of \(T^{G} \times H^{G} \times W^{G}\), and the number of channels \(C\) in LIVT, leveraged in the original \textbf{EvEnhancer}. As shown in Table~\ref{tab:Ab_Hyperparameter}, selecting an insufficient number of event segments, local grid size, or number of channels can lead to a degradation in reconstruction quality. Conversely, an overly large number of each can significantly decrease the model efficiency. Considering the balance between performance and complexity, we finally implement \textbf{EvEnhancerPlus} using the settings in the last two rows to compare with other methods in this work.
\begin{table}[t]
    \centering
    \caption{Impact of hyperparameters including the number of event segments \(M\) in EASM, the local grid size of \(T^{G} \times H^{G} \times W^{G}\), and the number of channels \(C\) in LIVT.}
    \centering
\resizebox{\columnwidth}{!}{\begin{tabular}{ccc||cc|cc|c}
    \hline
    \multirow{2}{*}{$M$}
    & \multirow{2}{*}{Grid Size}
    & \multirow{2}{*}{$C$}
    & \multicolumn{2}{c|}{\textit{In-Dist.}}
    & \multicolumn{2}{c|}{\textit{OOD}}
    & Param.
    \\
    
    &
    & 
    & PSNR & TFLOPs
    & PSNR & TFLOPs
    & (M)
    \\
    \hline
    5
    & 3\(\times\)3\(\times\)3
    & 64
    & 32.32 & 6.742
    & 28.73 & 8.512
    & 6.55 \\
    9
    & 3\(\times\)3\(\times\)3
    & 64
    & 33.08 & 7.516
    & 29.46 & 8.852
    & 6.55 \\
    7
    & 1\(\times\)3\(\times\)3
    & 64
    & 33.15 & 4.609
    & 27.00 & 5.077
    & 6.25 \\
    7
    & 5\(\times\)3\(\times\)3
    & 64
    & 33.35 & 9.649
    & 29.60 & 12.29
    & 6.84 \\
    7
    & 3\(\times\)1\(\times\)1
    & 64
    & 32.99 & 3.769
    & 29.29 & 3.875
    & 6.16 \\
    7
    & 3\(\times\)5\(\times\)5
    & 64
    & 33.28 & 13.85
    & 29.52 & 18.30
    & 7.34 \\
    \rowcolor{gray!20}
    7
    & 3\(\times\)3\(\times\)3
    & 16
    & 33.01 & 4.003
    & 29.12 & 4.416
    & 5.81 \\
    \rowcolor{gray!20}
    7
    & 3\(\times\)3\(\times\)3
    & 64
    & 33.30 & 7.129
    & 29.54 & 8.682
    & 6.55 \\
    \hline
    \end{tabular}}
    \label{tab:Ab_Hyperparameter}
\end{table}

\subsubsection{EvEnhancerPlus}
\label{sec:ablation_improvement}

\begin{figure}[t]
\centering
\includegraphics[width=\columnwidth]{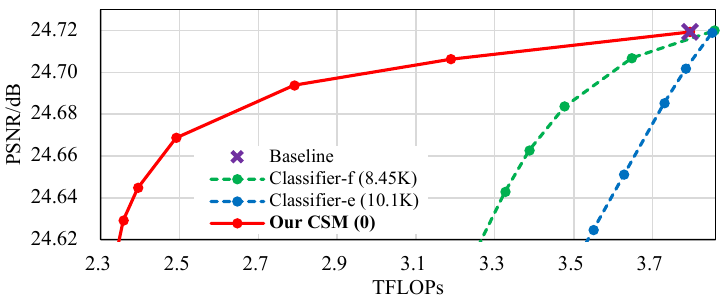}
\caption{Trade-off between PSNR and FLOPs among Classifier-f, Classifier-e, and our CSM on the BS-ERGB dataset~\cite{tulyakov2022time} at scale $t=6$, $s=4$. We use \textbf{EvEnhancerPlus} with a single complex upsampler as the baseline.}
\label{fig:curve}
\end{figure}

\textbf{Controllable Switch Mechanism (CSM).} The proposed CSM performs as a distributor that dynamically routes each pixel to computationally appropriate reconstruction pathways based on local event statistics. Here, we conduct ablations to study its effect. Specifically, we first implement two vanilla classifiers, defined as Classifier-f and Classifier-e, which estimate the reconstruction difficulty based on latent RGB frame features and event features, respectively. We employ the strategy presented in~\cite{jeong2024accelerating}, which utilizes a learnable MLP with pixel assignment supervision for this purpose. In Fig.~\ref{fig:distribution}, we apply three strategies to divide the pixels into two categories: simple and complex. It can be observed that our CSM is more effective in accurately distinguishing pixel characteristics. Moreover, since CSM allows for optimal computation allocation and adaptive routing pixels to different upsamplers, it achieves more computation savings than the baseline model and the other two classifiers, even producing SR images with high performance and better textures.

Then, to intuitively illustrate the advantages of CSM, Fig.~\ref{fig:curve} presents the PSNR–FLOPs curves for three competing strategies under varying classification probabilities (complex or simple, reflected by FLOPs). We can obtain three observations: 1) Since Classifier-f and Classifier-e apply trainable MLPs for pixel classification, they involve additional model parameters (8.45K \& 10.1K). In contrast, CSM remains entirely non-parametric; 2) CSM consistently maintains more favorable trade-offs between effectiveness and computational efficiency across all operating points; 3) CSM enables implementation at significantly lower computational budgets (2.3 TFLOPs \textit{vs.} 3.3 TFLOPs), revealing its preferable adaptability in practical scenarios.  

\begin{table}[t]
    \centering
    \caption{Ablation studies on the number of upsamplers $N$ in \textbf{EvEnhancerPlus}.}
    \centering
    \begin{tabular}{c||cc|cc|c}
    \hline
    \multirow{2}{*}{$N$}    
    & \multicolumn{2}{c|}{\textit{In-Dist.}}
    & \multicolumn{2}{c|}{\textit{OOD}}
    & Param.\\

    & PSNR
    & TFLOPs
    & PSNR
    & TFLOPs
    & (M)\\
    \hline
    1
    & 33.42
    & 6.956 (100\%)
    & 29.64
    & 8.655 (100\%)
    & 6.55\\
    \rowcolor{gray!20}
    2
    & 33.39
    & 6.125 (88.1\%)
    & 29.63
    & 7.771 (89.8\%)
    & 6.91\\
    3
    & 33.36
    & 5.702 (82.0\%)
    & 29.60
    & 7.093 (82.0\%)
    & 7.48 \\
    \hline
    \end{tabular}
    \label{tab:numUps}
\end{table}
\begin{figure}[t]
\centering
\includegraphics[width=\columnwidth]{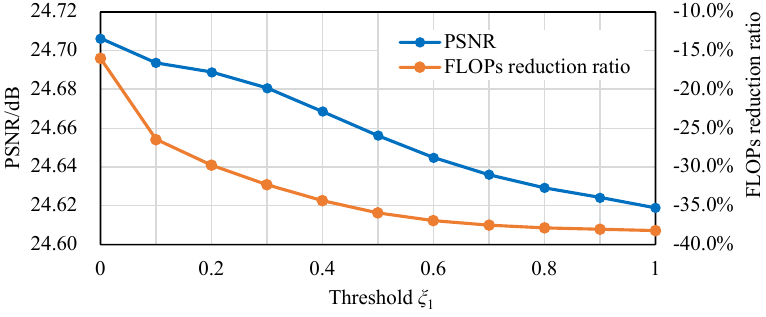}
\caption{Impact of the adjustable threshold $\xi_1$ on the performance and efficiency trade-offs in \textbf{EvEnhancerPlus}, evaluated on the BS-ERGB dataset~\cite{tulyakov2022time} at scale $t=6$, $s=4$ using PSNR (dB) and FLOPs reduction ratio (relative to the \textbf{EvEnhancerPlus} containing only the complex upsampler).}
\label{fig:curve_thershold}
\end{figure} 

Furthermore, we also investigate the impact of the number of upsamplers $N$ integrated within the CSM, where $N$ corresponds to the granularity allocation according to pixel reconstruction difficulty. As illustrated in Table~\ref{tab:numUps}, all multi-upsampler configurations ($N>1$) improve efficiency compared to the baseline ($N=1$). Besides, increasing $N$ from a binary scheme ($N=2$, simple/complex) to a three-tier setup ($N=3$, simple/medium/complex with LIVT backbones of 16, 32, and 64 channels, respectively) reduces computational costs but leads to more model parameters. Among these, $N=2$ achieves the best balance, which is used as the default setting in our experiments.

Finally, there are adjustable thresholds $\{\xi_n\}_{n=1}^{N-1}$ in CSM, which control pixel distribution (see Eq.~(\ref{eq:11})). In \textbf{EvEnhancerPlus}, for the implementation of two upsamplers, we have $\xi_1\in [0,1]$. To investigate the influence of $\xi_1$, as shown in Fig.~\ref{fig:curve_thershold}, we gradually change the values of $\xi_1$ to produce a PSNR-FLOP curve. One can see that the computational cost decreases with increasing $\xi_1$, since a larger portion of pixels is assigned to the simple upsampler. $\xi_1=1$ means that all pixels are addressed by a simple upsampler, thus involving minimal computations but at the degradation of reconstruction performance due to the overlook of high-difficulty pixels. Besides, since the reconstruction difficulty $R_{\tau}(x,y)$ of many pixels is 0, as shown in Fig.~\ref{fig:curve_thershold}, the model with $\xi_1=0$ exhibits the best trade-off. Moreover, due to the parameter-free nature of CSM, the threshold can be adjusted flexibly to further improve the efficiency in resource-constrained scenarios with a marginal performance drop.

\begin{table}[t]
    \centering
    \caption{Ablation studies for different training strategies on the GoPro dataset~\cite{nah2017deep} at \textit{OOD} scales ($t=12$, $s=6$). ``WS'' denotes weighted sum, and ``CDTS'' denotes our cross-derivative training strategy.}
    \centering
    \resizebox{\columnwidth}{!}{\begin{tabular}{c|c||cc|cc|cc}
    \hline
    Training
    & \multirow{2}{*}{Type}
    & \multicolumn{2}{c|}{Upsampler $\mathcal{U}_{0}$}
    & \multicolumn{2}{c|}{Upsampler $\mathcal{U}_{1}$}
    & \multicolumn{2}{c}{Overall}\\
    Scheme
    &
    & PSNR
    & SSIM
    & PSNR
    & SSIM
    & PSNR
    & SSIM
    \\
    \hline
    From Scratch
    & WS
    & 28.23 & 0.8183
    & 14.32 & 0.2960
    & 15.11 & 0.4010\\
    \hline
    \multirow{2}{*}{$\mathcal{U}_{0}\rightarrow\mathcal{U}_{1}$}
    & WS
    & 29.29 & 0.8515
    & 29.00 & 0.8425
    & 29.02 & 0.8431\\
    & \cellcolor{gray!20}\textbf{CDTS}
    & \cellcolor{gray!20}29.29 & \cellcolor{gray!20}0.8515
    & \cellcolor{gray!20}29.49 & \cellcolor{gray!20}0.8564
    & \cellcolor{gray!20}29.48 & \cellcolor{gray!20}0.8562\\
   \hline
    \multirow{2}{*}{$\mathcal{U}_{1}\rightarrow\mathcal{U}_{0}$}
    & WS
    & 28.58 & 0.8291
    & 29.64 & 0.8595
    & 29.62 & 0.8585\\
    & \cellcolor{gray!20} \textbf{CDTS}
    & \cellcolor{gray!20}29.10 & \cellcolor{gray!20}0.8455
    & \cellcolor{gray!20}29.64 & \cellcolor{gray!20}0.8595
    & \cellcolor{gray!20}\textbf{29.63} & \cellcolor{gray!20}\textbf{0.8591}\\
    \hline
    \end{tabular}}
    \label{tab:Ab_Strategy}
\end{table}

\textbf{Cross-Derivative Training Strategy.} Given that \textbf{EvEnhancerPlus} incorporates multiple upsamplers equipped by the CSM, we introduce a cross-derivative training strategy to ensure stable convergence. In Table~\ref{tab:Ab_Strategy}, we compare this approach against four alternative training strategies under identical settings. 1) Training the overall model with these upsamplers together from scratch, where the output is produced by a weighted sum (abbreviated as WS) based on the reconstruction difficulty estimated by CSM: 
$\hat{I}_{\tau}^{\textit{HR}}(x,y)=(1-\alpha)\mathcal{U}_{0}(\hat{\mathbf{F}}\text{; }(\tau,x,y))+\alpha \mathcal{U}_{1}(\hat{\mathbf{F}}\text{; }(\tau,x,y))$, where $\alpha$ is the reconstruction difficulty $R_{\tau}(x,y)$.
However, we observe a severe training bias under this setting, \emph{i.e.}, the simple upsampler $\mathcal{U}_{0}$ tends to dominate the learning process, while the complex one $\mathcal{U}_{1}$ fails to converge effectively, ultimately leading to unsatisfactory results. 2) Individually training different upsamplers in multiple stages with the sequence of $\mathcal{U}_0\rightarrow\mathcal{U}_1$ and generating the output also via a weighted sum. We can see that it acquires notable gains, particularly enhancing the reconstruction capability of the complex upsampler $\mathcal{U}_1$. 3) When we apply the proposed strategy (abbreviated as CDTS) on 2), since $\mathcal{U}_1$ can be initialized with prior knowledge inherited from the pre-trained
derivative network with $\mathcal{U}_0$, thus the performance improves. 4) Conducting individual training with the sequence of $\mathcal{U}_1\rightarrow\mathcal{U}_0$. We can see that the model is more conducive to recovering the pixels with high difficulty, showing better overall results than $\mathcal{U}_0\rightarrow\mathcal{U}_1$. This occurs because complex upsamplers play a decisive role in reconstructing challenging pixels, whose performance ultimately determines the overall quality ceiling of the final network. The $\mathcal{U}_1\rightarrow\mathcal{U}_0$ training sequence better preserves the representational capacity of $\mathcal{U}_1$, ensuring its superior reconstruction capability for difficult regions is maintained throughout the optimization process. Consequently, by using CDTS (the last row), the model achieves the best, which is employed in our final \textbf{EvEnhancerPlus}. These comparisons demonstrate the effectiveness of our cross-derivative training strategy. 

\section{Conclusion}
In this work, we present \textbf{EvEnhancer}, a novel approach for continuous space-time video super-resolution (C-STVSR) that effectively integrates the complementary advantages of event streams and video frames. By introducing the event-adapted synthesis module (EASM) and the local implicit video transformer (LIVT), our method enables accurate modeling of long-term motion trajectories and effective learning of continuous video representations. To further enhance computational efficiency, we propose \textbf{EvEnhancerPlus} to employ a parameter-free controllable switching mechanism (CSM) grounded in a theoretical approximation between events and frame changes. A cross-derivative training strategy stabilizes the convergence of \textbf{EvEnhancerPlus}. Extensive experiments on both synthetic and real-world datasets demonstrate that \textbf{EvEnhancerPlus} exhibits strong robustness and generalization with more pleasant efficiency.





\ifCLASSOPTIONcaptionsoff
  \newpage
\fi

\ifCLASSOPTIONcaptionsoff
\newpage
\fi

\bibliographystyle{IEEEtran}
\bibliography{main}

\end{document}